\documentclass[superscriptaddress,aps,onecolumn,12pt,floatfix,altaffilletter,showpacs,showkeys,notitlepage,nofootinbib]{revtex4-2}

\pdfoutput=1

\usepackage[utf8]{inputenc}
\usepackage[dvipdf,dvips,dvipdfmx]{graphicx}
\usepackage[colorlinks=true,citecolor=blue,linkcolor=blue]{hyperref}
\usepackage{amsmath}
\usepackage{amssymb}
\usepackage{soul,xcolor}
\usepackage{epsfig}
\usepackage{epstopdf}
\usepackage{url}
\usepackage{float}
\usepackage{slashed}
\usepackage{cancel}
\usepackage{textcomp}
\usepackage{calc}
\usepackage{adjustbox}
\usepackage{mathtools}
\usepackage{gauss}
\usepackage{physics}
\usepackage{gensymb}
\usepackage{bbm}
\usepackage{esint}
\usepackage{tikz-feynman}
\usepackage{bbold}
\usepackage{tensor}
\usepackage{kbordermatrix}
	\renewcommand{\kbldelim}{(}
	\renewcommand{\kbrdelim}{)}

\usepackage{indentfirst}
\usepackage[font=small,labelfont=bf]{caption}
\usepackage{wrapfig}
\usepackage{caption}
\usepackage{comment}
\usepackage{pgfplots}
	\pgfplotsset{width=5cm,compat=1.9}
\usepackage{bm}
\usepackage{fancyhdr}
\setstcolor{red}

\allowdisplaybreaks

\catcode`\@=11
\def\lsim{\mathrel{\mathpalette\@versim<}}
\def\gsim{\mathrel{\mathpalette\@versim>}}
\def\@versim#1#2{\vcenter{\offinterlineskip
\ialign{$\m@th#1\hfil##\hfil$\crcr#2\crcr\sim\crcr } }}
\catcode`\@=12

\newcommand{\Array}[1]{\begin{pmatrix} #1 \end{pmatrix}}
\newcommand{\Lag}{\mathcal{L}}
\newcommand{\Ham}{\mathcal{H}}
\newcommand{\Ord}{\mathcal{O}}

\newcommand{\Chg}{\mathcal{Q}}

\newcommand{\SM}{S}
\newcommand{\FS}{\mathcal{V}}
\newcommand{\FSp}{\mathcal{V}_{\text{phys}}}
\newcommand{\evltd}[1]{\Big\rvert_{#1}}
\newcommand{\met}{g_{\alpha\beta}}
\newcommand{\imet}{g^{\alpha\beta}}
\newcommand{\mmet}{\eta_{\alpha\beta}}
\newcommand{\immet}{\eta^{\alpha\beta}}
\newcommand{\grav}{h_{\alpha\beta}}
\newcommand{\igrav}{h^{\alpha\beta}}
\newcommand{\Hab}{H_{\alpha\beta}}
\newcommand{\iHab}{H^{\alpha\beta}}
\newcommand{\gdet}{\sqrt{-g}}

\newcommand{\trans}{\enskip\rightarrow\enskip}

\newcommand{\qwhere}{\quad\text{where}\quad}
\newcommand{\qcom}{\,,\quad}
\newcommand{\qper}{\,.}

\newcommand{\nreturn}{\nonumber \\[0.3em]}
\newcommand{\return}{\\[0.3em]}

\newcommand{\bc}{\begin{color}{red}}
\newcommand{\ec}{\end{color}}

\newcommand{\Com}[2]{[#1,#2]}
\newcommand{\bigPB}[2]{\big\{#1,#2\big\}}
\newcommand{\bigCom}[2]{\big[#1,#2\big]}
\newcommand{\BigPB}[2]{\Big\{#1,#2\Big\}}
\newcommand{\BigCom}[2]{\Big[#1,#2\Big]}
\newcommand{\sml}[1]{\mbox{\normalsize $#1$}}

\newcommand{\osc}[2]{#1_{\,{\raisebox{.5pt}{${\!}_{\scriptstyle #2}$}}}{}}
\newcommand{\oscd}[2]{#1^\dagger_{\,\raisebox{.5pt}{${\!}_{\scriptstyle #2}$}}{}}
\newcommand{\aop}[2]{a_{#1,#2}}
\newcommand{\polt}[1]{\varepsilon_{{\!}_{\scriptstyle #1}}{}}
\newcommand{\poltd}[1]{{\varepsilon}^*_{\raisebox{.5pt}{${\!}_{\scriptstyle #1}$}}{}}

\begin{document}

\title{Analysis of Unitarity in \\ Conformal Quantum Gravity}

\author{Jisuke \surname{Kubo}}
\email{kubo@mpi-hd.mpg.de}
\affiliation{Max-Planck-Institut f\"ur Kernphysik (MPIK), Saupfercheckweg 1, 69117 Heidelberg, Germany}
\affiliation{Department of Physics, University of Toyama, 3190 Gofuku, Toyama 930-8555, Japan}

\author{Jeffrey \surname{Kuntz}}
\email{jkuntz@mpi-hd.mpg.de}
\affiliation{Max-Planck-Institut f\"ur Kernphysik (MPIK), Saupfercheckweg 1, 69117 Heidelberg, Germany}

\date{\today}

\begin{abstract}
We perform a canonical quantization of Weyl's conformal gravity by means of the covariant operator formalism and investigate the unitarity of the resulting quantum theory. After reducing the originally fourth order theory to second order in time derivatives via the introduction of an auxiliary tensor field, we identify the full Fock space of quantum states under a BRST construction that includes Faddeev-Popov ghost fields corresponding to Weyl transformations. Using the Kugo-Ojima quartet mechanism, we identify the physical subspace of quantum states and find that the subspace containing the transverse spin-2 states comes equipped with an indefinite inner product metric and a one-particle Hamiltonian that possesses only a single eigenstate. We construct the LSZ reduction formula for the S-matrix in this spin-2 subspace and find that unitarity is violated in scattering events. The explicit way in which this violation occurs represents a new view on the ghost-problem in quadratic theories of quantum gravity.
\end{abstract}


\maketitle

\pagebreak

\section{Introduction}

It is well-known that if one employs standard quantization procedures to Einstein's General Relativity, the resulting theory does not represent a satisfactory quantum theory, simply due to the fact that the theory is power-counting non-renormalizable as a result of its massive coupling constant. Thus, modified theories of gravity possessing actions that are quadratic in powers of the Riemann tensor, and hence power-counting renormalizable, have long been a focal point for theorists in search of a consistent theory of quantum gravity. Indeed, Stelle famously showed that quadratic gravity is in fact renormalizable on quite general grounds in \cite{Stelle1977}. However, the very feature that makes quadratic gravity power-counting renormalizable, an action functional where four derivatives act on the metric, also generally leads to the presence of additional ghost degrees of freedom (as compared to the standard two derivative case) that can cause the classical Hamiltonian to become unbounded from below; an effect known as the Ostrogradsky instability \cite{Ostrogradsky1850, Woodard2015}. While there do exist certain higher order theories that do not possess this instability, namely, theories with a degenerate kinetic matrix \cite{Horndeski1974,Langlois2016}, there is still much active interest in the consistency of the more general class of theories of quantum gravity constructed from non-degenerate quadratic actions where the issue of unitarity violation at the quantum level is often referred to as the ``ghost problem''.

The general consensus regarding the ghost problem appears to be that even if a given theory exhibits an Ostrogradsky instability at the classical level, this does not necessarily mean that the quantum theory is also inconsistent. While it is true that a quantum Hamiltonian operator with negative eigenvalues implies a violation of unitarity at the quantum level, it is in principle possible to quantize a fourth order theory in such a way that the quantum Hamiltonian has only positive eigenvalues. In short, it may be possible to circumvent the ghost problem using an appropriate quantization procedure. Many interesting theories have been constructed in this spirit over the years; notable examples include the well-known work of Boulware, Horowitz, and Strominger \cite{Boulware1983}, the more modern works by Donoghue and Menezes in \cite{Donoghue2019,Donoghue2021} and Anselmi in \cite{Anselmi2018}, which are related to the original Lee-Wick models \cite{Lee1969a}, as well as the the Agravity prescription \cite{Salvio2016} and developments in $\mathcal{PT}$-symmetric QFT \cite{Bender2008,Mannheim2018}, among others.

In this work we present a novel view on the ghost problem in fourth order gravity by means of BRST quantization in the covariant operator formalism; a framework developed by Nakanishi, Kugo, Ojima, and their contemporaries in order to establish unitarity in Yang-Mills theories \cite{Kugo1979b,Nakanishi1990,Kugo}. This kind of analysis has been applied to General Relativity \cite{Kugo1978-2} and also recently to unimodular gravity \cite{Kugo2022}, though applying it to quadratic gravity is only possible after recasting the fourth order classical theory to second order by introducing auxiliary fields which, if integrated out, return the original fourth order action. We put our focus on conformal gravity, the particular quadratic gravitational theory that is not only globally scale invariant, but locally invariant under the full conformal group. We focus on conformal gravity specifically because of its higher level of symmetry, which as we will see, leads to a richer BRST structure as compared to the globally symmetric case. We have in fact performed all of the analogous computations in the body of this text for globally scale invariant quadratic gravity as well and found that the discussion on unitarity works in nearly the exact same fashion, with the key differences being an additional two physical scalar modes and a simpler Faddeev-Popov ghost system in the globally invariant case. All the important differences between the globally and locally symmetric cases will be discussed in the conclusion. 

The outline of this work goes as follows. In Section \ref{sec:classical} we establish the classical theory of second order conformal gravity (SOCG) that serves as our starting point, while Section \ref{sec:quant} is devoted to the quantization process. There we derive the gauge-fixed BRST-invariant SOCG total action at the non-linear level, identify the free part of action by linearizing the metric around flat space, and appeal to the LSZ formalism \cite{Lehmann1955} in order to decompose each of the fields in terms of plane wave functions and quantum oscillators. The (anti)commutation relations among these oscillators define the quantum theory and allows us to proceed to Section \ref{sec:unitarity} where we analyze the unitarity of quantum SOCG. This discussion is divided into two parts; the first going along standard lines with an application of BRST cohomology to demonstrate how the Kugo-Ojima quartet mechanism identifies a physical subspace of the full Fock space which contains only transverse spin-1 and spin-2 states. The next part of the unitarity discussion revolves around this physical space, in particular the spin-2 subspace of it which is found to possess an indefinite inner product. After discussing this interesting feature and how it affects the eigenstates of the quantum Hamiltonian operator, we identify how the ghost problem manifests itself in this covariant operator formalism when we encounter a violation of unitarity in scattering events. Summaries of the important formalisms employed here and supplemental calculations may be found in the appendices. 

Throughout this work we use the metric signature $(-1,1,1,1)$ and the Riemann tensor sign $\tensor{R}{_\alpha_\beta_\gamma^\delta}=-\partial_\alpha\tensor{\Gamma}{^\delta_\beta_\gamma}+\cdots$. We also note that we have made extensive use of the xAct suite of packages \cite{Martin-Garcia2008,Brizuela2009} for Wolfram Mathematica, as well as particular the xTras \cite{Nutma2014a} and FieldsX \cite{Frob2020} supplements, which are invaluable for long calculations involving many tensor indices as were necessary here.

\section{Classical conformal gravity at second order} \label{sec:classical}

Assuming a metric-compatible and torsion-free affine connection, the most general action for globally scale-invariant quadratic gravity is given by a sum of the squares of the Riemann tensor, Ricci tensor, and Ricci scalar,
\begin{align} 
S_\text{QG} = \int\dd^4x\gdet\Big(aR_{\alpha\beta\gamma\delta}R^{\alpha\beta\gamma\delta} + bR_{\alpha\beta}R^{\alpha\beta} + cR^2\Big) \,,
\end{align}
where $a$, $b$, and $c$ are arbitrary dimensionless constants. This action may be simplified by eliminating the Riemann square using the Gauß-Bonnet invariant
\begin{align}
\mathcal{G} = R_{\alpha\beta\gamma\delta}R^{\alpha\beta\gamma\delta} - 4R_{\alpha\beta}R^{\alpha\beta} + R^2 \,,
\end{align}
which is a total derivative i.e.\ a boundary term that may be set to zero at the level of the action \cite{Alvarez-Gaume2016}. One may also rewrite the square of the Ricci tensor in terms of the Weyl tensor as
\begin{align}
R_{\alpha\beta}R^{\alpha\beta} = \frac{1}{2}C_{\alpha\beta\gamma\delta}C^{\alpha\beta\gamma\delta} + \frac{1}{3}R^2 + (\text{boundary terms}) \qcom
\end{align}
which, after eliminating one arbitrary constant and redefining the other two in terms of the new constants $\alpha_g$ and $\beta$, leaves us with the following convenient form for the action of quadratic gravity.
\begin{align} \label{SQG}
S_\text{QG} = \frac{2}{\alpha_g^2}\int\dd^4x\gdet\Big(C_{\alpha\beta\gamma\delta}C^{\alpha\beta\gamma\delta} + \beta R^2\Big)
\end{align}
This action is invariant under infinitesimal local diffeomorphisms of the metric
\begin{align} \label{gdiff}
\delta_\xi\met = \alpha_g\Lag_\xi\met = \alpha_g\big(\nabla_\alpha\xi_\beta + \nabla_\beta\xi_\alpha\big) \,,
\end{align}
where $\Lag_\xi$ is the Lie derivative in the direction of the arbitrary vector $\xi^\alpha(x)$. Note that we have also included, without loss of generality, a factor of $\alpha_g$ into this transformation rule for later convenience as this constant will end up serving a dual role as a perturbation parameter. (\ref{SQG}) is also invariant under the global scale transformations $\met \rightarrow c\,g_{\alpha\beta}$ where $c$ is a constant.

If we set $\beta=0$ in (\ref{SQG}), then we arrive at the action describing Weyl's conformal gravity.
\begin{align} \label{SCG}
S_\text{CG} = \frac{2}{\alpha_g^2}\int\dd^4x\gdet\,C_{\alpha\beta\gamma\delta}C^{\alpha\beta\gamma\delta}
\end{align}
This action is invariant under the diffeomorphisms (\ref{gdiff}) as well as local conformal transformations that may be expressed as the infinitesimal Weyl transformations
\begin{align} \label{gconf}
\delta_\omega\met = \alpha_g\,\omega\met \,,
\end{align}
where $\omega(x)$ is an arbitrary scalar function.

Generally speaking, it is difficult to formulate any fourth order quantum theory from the operator perspective. However, one may get around this difficulty by starting with a second order theory that is classically equivalent to the fourth order theory of interest. We thus consider the action
\begin{align} \label{SH}
S_\text{H} = \int\dd^4x\gdet\left(\!\!-\frac{2}{\alpha_g}G_{\alpha\beta}\iHab - \frac{1}{4}\left(\Hab\iHab - \tensor{H}{_\alpha^\alpha}\tensor{H}{_\beta^\beta}\right)\right) \,,
\end{align}
where $G_{\alpha\beta}=R_{\alpha\beta}-\frac{1}{2}\met R$ is the Einstein tensor and $\Hab(x)$ is an auxiliary symmetric tensor field. Integrating this auxiliary field out of the action using its equation of motion (EOM)
\begin{align}
\frac{4}{\alpha_g}G_{\alpha\beta} + \Hab -\met\tensor{H}{_\gamma^\gamma} = 0 \,,
\end{align}
returns the original fourth order action for conformal gravity (\ref{SCG}) as desired. One may also introduce an additional auxiliary scalar $\chi(x)$ in order to establish a second order theory that is classically equivalent to quadratic gravity by adding the following action to (\ref{SH}).
\begin{align} \label{Schi}
S_\chi = \int\dd^4x\gdet\left(\!\!-\frac{2}{\alpha_g}R\chi - \frac{1}{2\beta}\chi^2\right)
\end{align}
Integrating $\Hab$ and $\chi$ from $S_\text{H}+S_\chi$ returns (\ref{SQG}), however, as discussed in the introduction, we will ignore these additional $S_\chi$ terms and focus on the specific case of conformal quadratic gravity from here on out. Due to the increased level of gauge symmetry in the conformal case, it is straightforward to extrapolate the discussion on unitarity from the conformal theory to the general quadratic theory, while much of the interesting BRST structure is lost if one tries to go the other direction.

Before proceeding with the quantization process, it is beneficial to perform an additional re-shuffling of the DOFs by introducing a Stückelberg vector field $A_\alpha(x)$ via
\begin{align}
\Hab \trans \Hab + \nabla_\alpha A_\beta + \nabla_\beta A_\alpha \,,
\end{align}
which, applied to (\ref{SH}), returns the action for second order conformal gravity (SOCG) that will serve as our classical starting point.
\begin{align} \label{SSOCG}
S_\text{SOCG} = &\int\dd^4x\gdet\bigg(\!\!-\frac{2}{\alpha_g}G_{\alpha\beta}\iHab - \frac{1}{4}\left(\Hab\iHab - \tensor{H}{_\alpha^\alpha}\tensor{H}{_\beta^\beta}\right) \nreturn
&-\frac{1}{4}F_{\alpha\beta}F^{\alpha\beta} + R_{\alpha\beta}A^\alpha A^\beta - \Hab\nabla^\beta A^\alpha + \tensor{H}{_\alpha^\alpha}\nabla_\beta A^\beta\bigg)
\end{align}
Here, $F_{\alpha\beta} = \nabla_\alpha A_\beta - \nabla_\beta A_\alpha = \partial_\alpha A_\beta - \partial_\beta A_\alpha$ as usual. $S_\text{SOCG}$ is invariant under the same diffeomorphisms (\ref{gdiff}), which act on $\Hab$ and $A_\alpha$ in the standard way,
\begin{align}
&\delta_\xi \Hab = \alpha_g\Lag_\xi \Hab = \alpha_g\big(H_{\alpha\gamma}\nabla_\beta\xi^\gamma + H_{\beta\gamma}\nabla_\alpha\xi^\gamma + \xi^\gamma\nabla_\gamma \Hab\big) \label{Hdiff} \\[0.4em]
&\delta_\xi A_{\alpha} = \alpha_g\Lag_\xi A_{\alpha} = \alpha_g\big(A_{\beta}\nabla_\alpha\xi^\beta + \xi^\beta\nabla_\beta A_{\alpha}\big) \label{Adiff} \,,
\end{align}
as well as the Weyl transformations (\ref{gconf}) with the new fields transforming as
\begin{align}
&\delta_\omega \Hab = 4\nabla_\beta\nabla_\alpha\omega + \alpha_g\big(A_\alpha\nabla_\beta\omega + A_\beta\nabla_\alpha\omega - \met A_\gamma\nabla^\gamma\omega\big) \label{Hconf} \\[0.4em]
&\delta_\omega A_{\alpha} = 0 \label{Aconf} \,.
\end{align}
There is also an additional diffeomorphism-style gauge symmetry present in (\ref{SSOCG}), that comes about as part of the Stückelberg procedure, which compensates for the additional degrees of freedom in $A_\alpha$. This new symmetry acts on each of the fields as
\begin{align}
&\delta_\zeta\met = 0 &\delta_\zeta \Hab = \nabla_\alpha\zeta_\beta + \nabla_\beta\zeta_\alpha &&\delta_\zeta A_{\alpha} = -\zeta_\alpha \,,
\end{align}
where $\zeta^\alpha(x)$ is an arbitrary vector field. 

The Stückelberg procedure performed here, the introduction of a new vector field and the associated compensating gauge symmetry, is equivalent to a change of phase space variables in the Hamiltonian picture. This change of variables yields a system of purely first class constraints whereas one finds that some second class constraints are present if $A_\alpha$ is not introduced. Indeed, we have performed a Dirac-Bergmann analysis of the action (\ref{SSOCG}) after linearizing around flat space and found a total of 48 phase space variables with 18 irreducible first class constraints and no second class constraints. Dirac's formula thus tells us that SOCG has a total of 6 physical degrees of freedom, just as expected from standard fourth order conformal gravity \cite{Riegert1984}. It is also important to point out that the classical equivalence between conformal gravity and SOCG is not effected by the Stückelberg procedure; one may still integrate $\Hab$ out of (\ref{SSOCG}) and arrive back at (\ref{SCG}).

\section{Quantization}  \label{sec:quant}

\subsection{The gauge-fixed total action}

To quantize SOCG, we follow a BRST procedure in the style of \cite{Nakanishi1990}, summarized in Appendix \ref{sec:BRSToverview}, and define sets of bosonic Nakanishi-Lautrup (NL) fields $\{b_\alpha(x),\,B_\alpha(x),\,B(x)\}$, fermionic ghosts $\{c^\alpha(x),\,C^\alpha(x),\,C(x)\}$, and fermionic anti-ghosts $\{\bar{c}_\alpha(x),\,\bar{C}_\alpha(x),\,\bar{C}(x)\}$, where the three fields in each of these sets correspond to diffeomorphism, ``Stückelberg-diffeomorphism'', and Weyl symmetry respectively. It is important to note that the anti-ghosts are independent of the regular ghosts and not related by Hermitian conjugation, indeed, all of these new fields are Hermitian. Terms involving these fields are added to the classical Lagrangian (\ref{SSOCG}) in order to gauge fix the theory and establish BRST symmetry; the NL fields act as auxiliary fields that enforce the chosen gauge conditions via their EOMs while the fermionic nature of the (anti)ghosts allows them to cancel the redundant DOFs that are inherently present as a result of the gauge symmetry. BRST transformations are generated by the BRST charge operator $\Chg$ and their specific forms are fixed so that the BRST transformation is nilpotent i.e.\ $\Chg^2=0$. It is also important to point out that the BRST algebra is graded in terms of a field's ``ghost number'', with the ghosts and anti-ghosts assigned ghost numbers of one and minus one respectively, while the bosons carry a ghost number of zero. All physically relevant quantities, including the total action, are then restricted to be of ghost number zero.

Under BRST symmetry, the three original fields in (\ref{SSOCG}) transform as a sum of their gauge transformations with the transformation parameters replaced by the associated ghost fields.
\begin{align}
\begin{aligned} \label{nlBRST1} 
&\delta_\epsilon\met = \epsilon\,\alpha_g\big(\nabla_\alpha c_\beta + \nabla_\beta c_\alpha + \met C\big) \return
&\delta_\epsilon \Hab = \epsilon\Big(\nabla_\alpha C_\beta + \nabla_\beta C_\alpha + 4\nabla_\beta\nabla_\alpha C \return
&\phantom{\delta_\epsilon \Hab = }+ \alpha_g\Big(\big(\nabla_\gamma\Hab + H_{\alpha\gamma}\nabla_\beta + H_{\beta\gamma}\nabla_\alpha\big)c^\gamma + \big(A_\alpha\nabla_\beta + A_\beta\nabla_\alpha - \met A_\gamma\nabla^\gamma\big)C\Big)\Big) \return
&\delta_\epsilon A_\alpha = \epsilon\Big(\!\!-C_\alpha + \alpha_g\big(\nabla_\beta A_{\alpha} + A_{\beta}\nabla_\alpha\big)c^\beta\Big)
\end{aligned}
\end{align}
Here, $\epsilon$ is a constant anti-commuting and anti-Hermitian parameter of the BRST transformation. The new fields also transform under this symmetry as follows. 
\begin{align}
\begin{aligned} \label{nlBRST2} 
&\delta_\epsilon b_\alpha = 0 &&\delta_\epsilon B_\alpha = 0 &&\delta_\epsilon B = 0 \return
&\delta_\epsilon c^\alpha = \epsilon\,\alpha_g\,c^\beta\partial_\beta c^\alpha \quad &&\delta_\epsilon C^\alpha = \epsilon\,\alpha_g\big(c^\beta\partial_\beta C^\alpha + C^\beta\partial_\beta c^\alpha + C^\alpha C\big) \quad &&\delta_\epsilon C = \epsilon\,\alpha_g\,c^\alpha\partial_\alpha C \return
&\delta_\epsilon \bar{c}_\alpha = \epsilon\,ib_\alpha &&\delta_\epsilon \bar{C}_\alpha = \epsilon iB_\alpha &&\delta_\epsilon \bar{C} = \epsilon iB
\end{aligned}
\end{align}

With the transformation rules established, we proceed by selecting convenient gauge fixing conditions. One term must be added for each gauge symmetry which is then accompanied by a ghost term whose form is fixed if BRST symmetry is to be established. For diffeomorphism invariance we select the de Donder condition $\partial_\beta(\gdet\imet)=0$ which corresponds to the gauge-fixing and ghost actions
\begin{align}
&S_{\text{gf}\xi} = \frac{1}{\alpha_g}\int\dd^4x\gdet\,b^\alpha g^{\beta\gamma}\bigg(\partial_\gamma\met - \frac{1}{2}\partial_\alpha g_{\beta\gamma}\bigg) \return
&S_{\text{FP}\xi} = -i\int\dd^4x\gdet\Big(c^\alpha\nabla^\beta + c^\beta\nabla^\alpha - \imet\big(c^\gamma\nabla_\gamma - C\big)\Big)\partial_\beta\bar{c}_\alpha \,.
\end{align}
The Stückelberg diffeomorphism is fixed with the analogous de Donder-style condition $\nabla_\beta\tensor{H}{_\alpha^\beta} - \frac{1}{2}\nabla_\alpha\tensor{H}{_\beta^\beta} = 0$, which yields the actions
\begin{align}
&S_{\text{gf}\zeta} = \int\dd^4x\gdet\,B_\alpha\bigg(\nabla_\beta\iHab - \frac{1}{2}\nabla^\alpha\tensor{H}{_\beta^\beta}\bigg) \return
&S_{\text{FP}\zeta} = i\int\dd^4x\gdet\,\bar{C}_\alpha\bigg(\nabla_\beta\nabla^\beta C^\alpha + C_\beta R^{\alpha\beta} + 2\big(\nabla^\beta\nabla^\alpha + R^{\alpha\beta}\big)\nabla_\beta C \nreturn
&\phantom{S_{\text{FP}\zeta} =}+ \alpha_g\Big(\big(\imet\big(\nabla_\gamma c^\gamma + c^\gamma\nabla_\gamma\big) - \nabla^\beta c^\alpha\big)\big(\nabla_\delta\tensor{H}{_\beta^\delta} - \frac{1}{2}\nabla_\beta\tensor{H}{_\delta^\delta}\big) \nreturn
&\phantom{S_{\text{FP}\zeta} =}+ \big(\iHab + \nabla^\beta A^\alpha + \imet\nabla_\gamma A^\gamma + A^\alpha\nabla^\beta + A^\beta\nabla^\alpha\big)\nabla_\beta C\Big)\bigg) \,.
\end{align}
Finally, we introduce the gauge-fixing and ghost actions corresponding to the conformal condition $\frac{1}{2}\tensor{H}{_\alpha^\alpha} + \nabla_\alpha A^\alpha + \frac{1}{2}B = 0$, which has been chosen to make the propagators appear as convenient as possible and ends up roughly corresponding to the Feynman gauge with respect to $A_\alpha$.
\begin{align}
&S_{\text{gf}\omega} = \frac{1}{8}\int\dd^4x\gdet\,B\big(2\tensor{H}{_\alpha^\alpha} + 4\nabla_\alpha A^\alpha + B\big) \return
&S_{\text{FP}\omega} = i\int\dd^4x\gdet\,\bar{C}\bigg(\nabla_\alpha\nabla^\alpha C \nreturn
&\phantom{S_{\text{FP}\omega} = }+ \frac{\alpha_g}{4}\Big(\big(\nabla_\alpha c^\alpha + c^\alpha\nabla_\alpha + C\big)\big(\tensor{H}{_\beta^\beta} + 2\nabla_\beta A^\beta\big) + B\Big(\frac{1}{2}\nabla_\alpha c^\alpha + C\Big)\Big)\bigg)
\end{align}
The sum of all of these actions gives us the total BRST action $S_\text{T}$.
\begin{align} \label{ST}
S_\text{T} = S_\text{SOCG} + S_{\text{gf}\xi} + S_{\text{gf}\zeta} + S_{\text{gf}\omega} + S_{\text{FP}\xi} + S_{\text{FP}\zeta} + S_{\text{FP}\omega}
\end{align}

\subsection{Linearization and propagators}

The focus of this work is to investigate the unitarity of conformal quantum gravity, so our next task is identify the free part of the action as this part is what determines whether the theory is unitary or not. To this end, we linearize the total action by writing the metric as graviton perturbations around the flat space Minkowski metric $\mmet$,
\begin{align}
\met \trans \mmet + \alpha_g\grav \,,
\end{align}
where we have assumed that $\alpha_g$ is small so that it may serve as a perturbation parameter. After performing this linearization, all indices are to be contracted with the background metric $\mmet$. The linearized total action may then be written as a sum of the free action $S_0$ and an interaction action $S_\text{int}$,
\begin{align}
S_\text{T}\evltd{\met \rightarrow \mmet + \alpha_g\grav} = S_0 + S_\text{int} \,,
\end{align}
where $S_\text{int}=\Ord(\alpha_g)$ and $S_0$ is given by the following.
\begin{align} \label{S0}
S_0 = &\int\dd^4x\bigg(4\iHab\mathcal{E}_{\alpha\beta\gamma\delta}h^{\gamma\delta} - \frac{1}{4}\left(\Hab\iHab - \tensor{H}{_\alpha^\alpha}\tensor{H}{_\beta^\beta}\right) \nreturn
&- \frac{1}{4}F_{\alpha\beta}F^{\alpha\beta} - \Hab\partial^\alpha A^\beta + \tensor{H}{_\alpha^\alpha}\partial_\beta A^\beta \nreturn
&+ b_\alpha\Big(\partial_\beta\igrav - \frac{1}{2}\partial^\alpha\tensor{H}{_\beta^\beta}\Big) + B_\alpha\Big(\partial_\beta \iHab - \frac{1}{2}\partial^\alpha\tensor{h}{_\beta^\beta}\Big) + \frac{1}{8}B\big(2\tensor{H}{_\alpha^\alpha} + 4\partial_\alpha A^\alpha + B\big) \nreturn
&+ i\Big(\bar{c}_\alpha\big(\Box\,c^\alpha - \partial^\alpha C\big) + \bar{C}_\alpha\big(\Box\,C^\alpha + 2\Box\partial^\alpha C\big) + \bar{C}\,\Box\,C\Big)\bigg)
\end{align}
Here, $\Box=\partial_\alpha\partial^\alpha$ is the d'Alembertian operator and $\mathcal{E}_{\alpha\beta\gamma\delta}$ is the Lichnerowicz operator i.e.\ the kinetic operator for linearized General Relativity,
\begin{gather}
\bigg(\frac{\gdet}{\kappa^2}\,R\bigg)\evltd{\met \rightarrow \mmet + \kappa\grav} = \igrav\mathcal{E}_{\alpha\beta\gamma\delta}h^{\gamma\delta} + \Ord(\kappa) \return
\mathcal{E}_{\alpha\beta\gamma\delta} = \frac{1}{4}\Big(\big(\delta_{\alpha\beta\gamma\delta} - \eta_{\alpha\beta}\eta_{\gamma\delta}\big)\Box - \mathcal{D}_{\alpha\beta\gamma\delta} + \eta_{\alpha\beta}\partial_\gamma\partial_\delta + \eta_{\gamma\delta}\partial_\alpha\partial_\beta\Big)
\end{gather}
where we have used the shorthand notations
\begin{gather}
\delta_{\alpha\beta\gamma\delta} \equiv \frac{1}{2}(\eta_{\alpha\gamma}\eta_{\beta\delta} + \eta_{\alpha\delta}\eta_{\beta\gamma}) \return
\mathcal{D}_{\alpha\beta\gamma\delta} \equiv \frac{1}{2}\Big(\eta_{\alpha\gamma}\partial_\beta\partial_\delta + \eta_{\alpha\delta}\partial_\beta\partial_\gamma + \eta_{\beta\gamma}\partial_\alpha\partial_\delta + \eta_{\beta\delta}\partial_\alpha\partial_\gamma\Big) \,.
\end{gather}
Crucially, BRST invariance is preserved by this linearization procedure and one finds that the free action (\ref{S0}) is invariant under 
\begin{align}\label{linBRST}
\begin{aligned} 
&\delta_{\epsilon}\grav = \epsilon\big(\partial_\alpha c_\beta + \partial_\beta c_\alpha + \mmet\,C\big) &&\delta_{\epsilon}\Hab = \epsilon\big(\partial_\alpha C_\beta + \partial_\beta C_\alpha + 4\partial_\alpha\partial_\beta C\big) &&\delta_{\epsilon}A_\alpha = -\epsilon\,C_\alpha \return
&\delta_{\epsilon}b_\alpha = 0 &&\delta_{\epsilon}B_\alpha = 0 &&\delta_{\epsilon}B = 0 \return 
&\delta_{\epsilon}c^\alpha = 0 &&\delta_{\epsilon}C^\alpha = 0 &&\delta_{\epsilon}C = 0 \return
&\delta_{\epsilon}\bar{c}_\alpha = \epsilon\,ib_\alpha &&\delta_{\epsilon}\bar{B}_\alpha = \epsilon\,iB_\alpha &&\delta_{\epsilon}\bar{C} = \epsilon\,iB
\end{aligned}
\end{align}
which are simply the linearized versions of the full transformation rules (\ref{nlBRST1},\,\ref{nlBRST2}) at zeroth order in $\alpha_g$.

The inverse propagator that results from the free action (\ref{S0}) is given by the Fourier transform of the Hessian of the total action with respect to the complete set of fields $\Phi^A$,
\begin{align} \label{iProp}
\Omega^{-1}_{AB}(p) = \int\dd^4x\frac{\delta^2 S_\text{T}}{\delta\Phi^A(x)\delta\Phi^B(y)}\,e^{-ip(x-y)} = \Array{
	\Omega^{-1}_\text{boson} &  0 \\
	0 & \Omega^{-1}_\text{ghost}}_{AB}
\end{align}
\begin{align}
&\Omega^{-1}_\text{boson} = \kbordermatrix{
	& h_{\mu\nu} & H_{\mu\nu} & A_\mu & b_\mu & B_\mu & B \\
	h_{\alpha\beta} & 0 & \sml{F^{(1)}_{\alpha\beta\mu\nu}} & 0 & \sml{F^{(2)}_{\alpha\beta\mu}} & 0 & 0 \\
	H_{\alpha\beta} &  & \sml{\frac{1}{2}(\eta_{\alpha\beta}\eta_{\mu\nu} - \delta_{\alpha\beta\mu\nu})} & \sml{F^{(2)}_{\alpha\beta\mu}} & 0 & \sml{F^{(2)}_{\alpha\beta\mu}} & \sml{\frac{1}{4}\eta_{\alpha\beta}} \\
	A_\alpha &  &  & \sml{p_\alpha p_\mu - \eta_{\alpha\mu}p^2} & 0 & 0 & \sml{-\frac{i}{2}p_\alpha} \\
	b_\alpha &  &  &  & 0 & 0 & 0 \\
	B_\alpha &  & \text{(h.c.)} &  &  & 0 & 0 \\
	B &  &  &  &  &  & \frac{1}{4}
	} \return
&\Omega^{-1}_\text{ghost} = \kbordermatrix{
	& c_\mu & C_\mu & C & \bar{c}_\mu & \bar{C}_\mu & \bar{C}  \\
	c_\alpha & 0 & 0 & 0 & \sml{-i\eta_{\alpha\mu}p^2} & 0 & 0 \\
	C_\alpha &  & 0 & 0 & 0 & \sml{-i\eta_{\alpha\mu}p^2} & 0 \\
	C &  &  & 0 & \sml{-p_\mu} & \sml{-2p_\mu p^2} & \sml{-ip^2} \\
	\bar{c}_\alpha &  &  &  & 0 & 0 & 0 \\
	\bar{C}_\alpha &  & \text{(h.c.)} &  &  & 0 & 0 \\
	\bar{C} &  &  &  &  &  & 0}
\end{align}
where
\begin{align}
&F^{(1)}_{\alpha\beta\mu\nu} = \big(\eta_{\alpha\beta}\eta_{\mu\nu} - \delta_{\alpha\beta\mu\nu}\big)p^2 - \mmet p_\mu p_\nu - \eta_{\gamma\delta}p_\alpha p_\beta \nreturn
&\phantom{F^{(1)}_{\alpha\beta\mu\nu} =}+ \frac{1}{2}\big(\eta_{\alpha\mu} p_\beta p_\nu + \eta_{\alpha\nu}p_\beta p_\mu + \eta_{\beta\mu}p_\alpha p_\nu + \eta_{\beta\nu}p_\alpha p_\mu\big) \return
&F^{(2)}_{\alpha\beta\mu} = \frac{i}{2}\big(\mmet p_\mu - \eta_{\alpha\mu}p_\beta - \eta_{\beta\mu}p_\alpha\big) \,.
\end{align}
The propagator $\Omega^{AB}(p)$ is then defined as the inverse of $\Omega^{-1}_{AB}(p)$ in the sense that
\begin{align}
\Omega^{-1}_{AB}\Omega^{BC} = \text{diag}\big(\tensor{\delta}{_\alpha_\beta^\gamma^\delta},\,\tensor{\delta}{_\alpha_\beta^\gamma^\delta},\,\tensor{\delta}{_\alpha^\gamma},\,\tensor{\delta}{_\alpha^\gamma},\,\tensor{\delta}{_\alpha^\gamma},\,1,\,\tensor{\delta}{_\alpha^\gamma},\,\tensor{\delta}{_\alpha^\gamma},\,1,\,\tensor{\delta}{_\alpha^\gamma},\,\tensor{\delta}{_\alpha^\gamma},\,1\big)\,,
\end{align}
and is given explicitly by
\begin{align} \label{Prop}
\Omega^{AB}(p) = \bra{0}T\Phi^A\Phi^B\ket{0} = \Array{
	 \Omega_\text{boson} &  0 \\
	 \mbox{\large $0$} & \Omega_\text{ghost}}^{AB}
\end{align}
\begin{align}
&\Omega_\text{boson} = \kbordermatrix{
	& h^{\gamma\delta} & H^{\gamma\delta} & A^\gamma & b^\gamma & B^\gamma & B \\
	h^{\mu\nu} & \sml{-\frac{\scalebox{.95}{$G^{\mu\nu\gamma\delta}$}}{2p^2}} & \sml{\scalebox{.95}{$G^{\mu\nu\gamma\delta}$}} & 0 & \sml{-\frac{i(\eta^{\nu\gamma}p^\mu + \eta^{\mu\gamma}p^\nu)}{p^2}} & 0 & \sml{-\frac{2p^\mu p^\nu}{p^4} - \frac{\eta^{\mu\nu}}{p^2}} \\
	H^{\mu\nu} &  & 0 & 0 & 0 & \sml{-\frac{i(\eta^{\nu\gamma}p^\mu + \eta^{\mu\gamma}p^\nu)}{p^2}} & 0 \\
	A^\mu &  &  & \sml{-\frac{\eta^{\mu\gamma}}{p^2}} & 0 & \sml{\frac{\eta^{\mu\gamma}}{p^2}} & \sml{-\frac{2ip^\mu}{p^2}} \\
	b^\mu &  &  &  & 0 & 0 & 0 \\
	B^\mu &  & \text{(h.c.)} &  &  & 0 & 0 \\
	B &  &  &  &  &  & 0
	}
\end{align}
\clearpage
\begin{align}
&\Omega_\text{ghost} = \kbordermatrix{
	& c^\gamma & C^\gamma & C & \bar{c}^\gamma & \bar{C}^\gamma & \bar{C} \\
	c^\mu & 0 & 0 & 0 & \sml{-\frac{i\eta^{\mu\gamma}}{p^2}} & 0 & \sml{-\frac{p^\mu}{p^4}} \\
	C^\mu &  & 0 & 0 & 0 & \sml{-\frac{i\eta^{\mu\gamma}}{p^2}} & \sml{-\frac{2p^\mu}{p^2}} \\
	C &  &  & 0 & 0 & 0 & \sml{-\frac{i}{p^2}} \\
	\bar{c}^\mu &  &  &  & 0 & 0 & 0 \\
	\bar{C}^\mu &  & \text{(h.c.)} &  &  & 0 & 0 \\
	\bar{C} &  &  &  &  &  & 0}
\end{align}
where
\begin{align}
&G^{\mu\nu\gamma\delta} = \frac{1}{p^2}\Big(\frac{1}{2}\eta^{\mu\nu}\eta^{\gamma\delta} - \delta^{\mu\nu\gamma\delta}\Big) + \frac{1}{2p^4}\big(\eta^{\mu\gamma}p^\nu p^\delta + \eta^{\mu\delta}p^\nu p^\gamma + \eta^{\nu\gamma}p^\mu p^\delta + \eta^{\nu\delta}p^\mu p^\gamma\big) \,.
\end{align}
With the propagator in hand, we may proceed with the quantization process by appealing to the LSZ formalism in order to establish asymptotic solutions to the equations of motion of our system.

\subsection{Asymptotic fields} \label{sec:AsymFields}

In the LSZ formalism, one treats the fields in a theory $\Phi(x)$ as Heisenberg fields i.e.\ as quantum fields with time-independent state vectors, and makes the assumption that at times $t=x^0\to\pm\infty$, the $\Phi(x)$ behave as a free fields that satisfy the free equations of motion\footnote{Strictly speaking, the limit $x^0\to\pm\infty$ should be  a weak limit. We also ignore effects such as wave function renormalization here since they will not affect the essence of the work that follows.} \cite{Nakanishi1990}.
\begin{align}
\Phi(x) \to 
\left\{\begin{array}{c}
	\Phi^\text{in}(x) \,, \quad x^0 \to -\infty \\ 
	\Phi^\text{out}(x) \,, \quad x^0 \to +\infty
\end{array}\right.
\end{align}
The formalism dictates that each asymptotic field may be decomposed as a sum of products of oscillators and plane wave functions as
\begin{align} \label{PhiDecomp}
\Phi^{\text{as}}(x) = \sum_{\bm p}\Big(\Phi_f^{\text{as}}({\bm p})f_{\bm p}(x) + \Phi_g^{\text{as}}({\bm p})g_{\bm p}(x) + \Phi_h^{\text{as}}({\bm p})h_{\bm p}(x) + \cdots + (\mbox{h.c.})\Big) \qcom
\end{align}
where ${\bm p}$ stands for the three-dimensional spatial part of the four-momentum $p^\alpha$. Here, the plane wave functions $f_{\bm p}(x)$, $g_{\bm p}(x)$, $h_{\bm p}(x)$, $\cdots$, are solutions to increasing powers of the d'Alembert equation (see Appendix \ref{sec:planewaves} for more on these functions).
\begin{align}
\Box\,f_{\bm p}(x) = \Box^2g_{\bm p}(x) = \Box^3h_{\bm p}(x) = \cdots = 0
\end{align}
The operator $\Phi_f^{\text{as}}({\bm p})$ in (\ref{PhiDecomp}) represents the fundamental simple-pole oscillator associated with the Heisenberg field $\Phi(x)$, where the superscript ``as'' stands for ``in'' or ``out'' depending on which limit is taken. The dipole, tripole, etc.\ oscillators (indicated with subscript $g$, $h$, ...) are not independent DOFs, but are rather functions of the fundamental oscillators and must be solved for using the equations of motion. One may also invert (\ref{PhiDecomp}) in order to define a fundamental oscillator in terms of its Heisenberg field using the orthogonality relations (\ref{ortho1}\,--\,\ref{ortho4}) which yields
\begin{align} \label{PhihatDef}
\Phi_f^{\text{as}}({\bm p}) = \lim_{x^0\to\pm\infty}\bigg[\,i\int\dd^3\bm x \Big(f_{\bm p}^*(x)\overset{\leftrightarrow}{\partial}_0 + g_{\bm p}^*(x)\overset{\leftrightarrow}{\partial}_0\Box + h_{\bm p}^*(x)\overset{\leftrightarrow}{\partial}_0\Box^2 + \cdots \Big)\Phi(x)\bigg] \,,
\end{align}
where we have used the shorthand $A\overset{\leftrightarrow}{\partial}_0B \equiv A\partial_0B-B\partial_0A$. These fundamental oscillators are products of annihilation(creation) operators and polarization tensors (which are non-trivial when $\Phi(x)$ corresponds to a field carrying space-time indices), as will be elucidated in the coming sections.

It is also important to note that the plane wave functions are normalized in a finite volume $V$, though we will also need to consider the continuum limit $V\to\infty$ when defining (anti)commutation relations among the oscillators so that sums over $\bm p$ may be exchanged for integrals according to
\begin{align} \label{continuum}
\lim_{V\to\infty} \left(\frac{1}{V}\sum_{\bm p}\right) = \int\frac{\dd^3{\bm p}}{(2\pi)^3} \qper
\end{align}
This is an important consideration since it implies that all (anti)commutation relations derived in a specific Lorentz frame will remain valid in any other frame. To see this one need only rescale each operator according to
\begin{align}
\Phi(\bm p) \to (2\pi)^{3/2}V^{-1/2}\Phi(\bm p) \qper
\end{align}
The rescaled operators are then found to be Lorentz scalars in the continuum limit (\ref{continuum}) i.e.\ $\sqrt{E'}\phi'(\bm p) \to \sqrt{E}\phi(\bm p)$, allowing us to recover the Lorentz-invariant statement $E'\delta^3({\bm p'} - {\bm q'}) = E\delta^3({\bm p} - {\bm q})$, where $E=|\bm p|$ as usual.

Now, turning back to the theory at hand, the equations of motion obtained from the total action (\ref{ST}) in the bosonic sector are given by
\begin{align} \label{BoEOMs}
\begin{aligned}
&\mathcal{E}_{\alpha\beta\gamma\delta}h^{\gamma\delta} - \frac{1}{8}\bigg(H_{\alpha\beta} + \partial_\alpha\big(A_\beta + B_\beta\big) + \partial_\beta\big(A_\alpha + B_\alpha \big) \return
&- \eta_{\alpha\beta}\Big(\tensor{H}{_\gamma^\gamma} + \partial_\gamma\big(2A^\gamma + B^\gamma\big) + \frac{1}{2}B\Big)\bigg) = 0 \return
&\mathcal{E}_{\alpha\beta\gamma\delta}H^{\gamma\delta} - \frac{1}{8}\Big(\partial_\alpha\,b_\beta + \partial_\beta\,b_\alpha - \eta_{\alpha\beta}\partial_\gamma\,b^\gamma\Big) = 0 \return
&\partial^\beta F_{\alpha\beta} - \partial^\beta H_{\alpha\beta} + \partial_\alpha\tensor{H}{_\beta^\beta} + \frac{1}{2}\partial_\alpha B = 0  \return
&\partial^\beta h_{\alpha\beta} - \frac{1}{2}\partial_\alpha \tensor{h}{_\beta^\beta} = 0 \return
&\partial^\beta H_{\alpha\beta} - \frac{1}{2}\partial_\alpha \tensor{H}{_\beta^\beta} = 0 \return
&\tensor{H}{_\alpha^\alpha} + 2\partial_\alpha A^\alpha + B = 0 \,.
\end{aligned}
\end{align}
This full set of EOMs is difficult to solve all at once, so it is convenient to first combine them and their traces so as to isolate the d'Alembertian of each field.
\begin{align} \label{sBoEOMs}
\begin{aligned}
&\Box\,h_{\alpha\beta} - \frac{1}{2}\Big(H_{\alpha\beta} + \partial_\alpha\big(A_\beta + B_\beta\big) + \partial_\beta\big(A_\alpha + B_\alpha\big)\Big) = 0 \return
&\Box\,H_{\alpha\beta} - \frac{1}{2}\Big(\partial_\alpha b_\beta + \partial_\beta b_\alpha\Big) = 0 \return
&\Box\,A_\alpha = 0
\end{aligned}
\end{align}

Now we must solve each of these equations by decomposing the Heisenberg fields as in (\ref{PhiDecomp}). To avoid clutter, we assume an implicit sum/integral over $\bm p$ where appropriate and drop the ``as'' designation until it becomes important again. The last equation above indicates that $A_\alpha(x)$ contains only a simple-pole, as expected from the propagator,
\begin{align} \label{Adecomp}
A^{\alpha}(x) = A_f^{\alpha}({\bm p})f_{\bm p}(x) + (\mbox{h.c.}) \,,
\end{align}
however, the tensor fields necessarily contain multi-pole contributions. We thus make the following ansätze for each of them based on the $p^{-2n}$ nature of their propagators ($n=$ number of poles required),
\begin{align}
&h^{\alpha\beta}(x) = h_f^{\alpha\beta}({\bm p})f_{\bm p}(x) + h_{g}^{\alpha\beta}({\bm p})g_{\bm p}(x) + h_{h}^{\alpha\beta}({\bm p})h_{\bm p}(x) + (\mbox{h.c.}) \label{hdecomp} \return
&H^{\alpha\beta}(x) = H_f^{\alpha\beta}({\bm p})f_{\bm p}(x) + H_{g}^{\alpha\beta}({\bm p})g_{\bm p}(x) + (\mbox{h.c.}) \,, \label{Hdecomp}
\end{align}
and find that the simplified EOMs (\ref{sBoEOMs}) may be solved with
\begin{align}
&h_{g}^{\alpha\beta}({\bm p})g_{\bm p}(x) = \frac{1}{2}\Big(H_f^{\alpha\beta}({\bm p}) + ip^\alpha A_f^{\beta}({\bm p}) + ip^\beta A_f^{\alpha}({\bm p}) + B_f^{\alpha}({\bm p})\partial^\beta + B_f^{\beta}({\bm p})\partial^\alpha\Big)g_{\bm p}(x) \return
&h_{h}^{\alpha\beta}({\bm p})h_{\bm p}(x) = \frac{1}{4}\Big(b_f^{\alpha}({\bm p})\partial^\beta + b_f^{\beta}({\bm p})\partial^\alpha\Big)h_{\bm p}(x) \return
&H_{g}^{\alpha\beta}({\bm p})g_{\bm p}(x) = \frac{1}{2}\Big(b_f^{\alpha}({\bm p})\partial^\beta + b_f^{\beta}({\bm p})\partial^\alpha\Big)g_{\bm p}(x) \,.
\end{align}
To solve the full set of bosonic EOMs (\ref{BoEOMs}), we must also decompose the NL fields as
\begin{align}
&b^\alpha(x) = b_f^\alpha({\bm p})f_{\bm p}(x) + (\mbox{h.c.}) \quad \return
&B^\alpha(x) = B_f^\alpha({\bm p})f_{\bm p}(x) + (\mbox{h.c.})\quad \return
&B(x) = B_f({\bm p})f_{\bm p}(x) + B_g({\bm p})g_{\bm p}(x) + (\mbox{h.c.})
\end{align}
and take advantage of the fact that the last three equations in (\ref{BoEOMs}), the NL EOMs i.e.\ the gauge fixing conditions, require that the longitudinal parts of $h_f^{\alpha\beta}$, $H_f^{\alpha\beta}$, and $A_f^{\alpha}$ are not independent from the other fields. We thus find that (\ref{BoEOMs}) may be solved with the decompositions (\ref{Adecomp}\,--\,\ref{Hdecomp}) along with
\begin{align}
B_g({\bm p})g_{\bm p}(x) = b_f^\alpha({\bm p})\big(ip_\alpha - 2\partial_\alpha\big)g_{\bm p}(x) \,,
\end{align}
and the gauge constraints
\begin{align}
&p_\beta h_f^{\alpha\beta}(\bm p) = \frac{1}{2}p^\alpha\tensor{{h_f}}{_\beta^\beta}(\bm p) + \frac{i}{4}\big(A_f^{\alpha}(\bm p) + 2B_f^{\alpha}(\bm p)\big) \nreturn
&\phantom{p_\beta h_f^{\alpha\beta}(\bm p) =} - \frac{1}{16E}\eta_{0\beta}\Big(4H_f^{\alpha\beta}(\bm p) + 4ip^\alpha A_f^\beta(\bm p) - \frac{1}{E}\tensor{\eta}{_0^\alpha}b_f^\beta(\bm p) + 2\eta^{\alpha\beta}B_f(\bm p)\Big) \label{transh} \return
&p_\beta H_f^{\alpha\beta}(\bm p) = \frac{1}{2}\Big(p^\alpha\tensor{{H_f}}{_\beta^\beta}(\bm p) + ib_f^{\alpha}(\bm p)\Big) \label{transH} \return
&p_\alpha A_f^{\alpha}(\bm p) = \frac{i}{2}\Big(\tensor{{H_f}}{_\alpha^\alpha}(\bm p) - B_f(\bm p)\Big) + \frac{1}{4E}\eta_{0\alpha}b_f^\alpha(\bm p) \label{transA} \,.
\end{align}
These relations come as a result of the particular gauge that we have chosen and are crucial for properly restricting the degrees of freedom. Indeed, with these restrictions one finds $h_{\alpha\beta}$, $H_{\alpha\beta}$, and $A_\alpha$ carry only 6, 6, and 3 independent degrees of freedom respectively. It is then straightforward to count a total of 24 bosonic degrees of freedom after including the 9 NL fields, 18 of which will cancel with the ghosts and anti-ghosts, leaving 6 physical degrees of freedom as expected. 

The ghost decompositions follow in the same manner as the bosons, though the process is simpler thanks to a lack of gauge constraints in that sector. Their EOMs are given by
\begin{align}
\begin{aligned}
&\Box\,c^{\alpha} - \partial^\alpha C = 0 &&\Box\,\bar{c}_{\alpha} = 0 \return
&\Box\,C^{\alpha} + 2\Box\partial^\alpha C = 0 \qquad\qquad\qquad &&\Box\,\bar{C}_{\alpha} = 0 \return
&\Box\,C = 0 &&\Box\,\bar{C} + \partial_\alpha\bar{c}^\alpha - 2\Box\partial_\alpha\bar{C}^\alpha  = 0 \,,
\end{aligned}
\end{align}
which indicate that the ghosts must decompose as
\begin{align}
&c^\alpha(x) = c_f^{\,\alpha}({\bm p})f_{\bm p}(x) + c_{g}^{\,\alpha}({\bm p})g_{\bm p}(x) + (\mbox{h.c.}) \qquad &&\bar{c}^{\,\alpha}(x) = \bar{c}_f^{\,\alpha}({\bm p})f_{\bm p}(x) + (\mbox{h.c.}) \nreturn
&C^\alpha(x) = C_f^\alpha({\bm p})f_{\bm p}(x) + (\mbox{h.c.}) &&\bar{C}^\alpha(x) = \bar{C}_f^\alpha({\bm p})f_{\bm p}(x) + (\mbox{h.c.}) \return
&C(x) = C_f({\bm p})f_{\bm p}(x) + (\mbox{h.c.}) &&\bar{C}(x) = \bar{C}_f({\bm p})f_{\bm p}(x) + \bar{C}_{g}({\bm p})g_{\bm p}(x) + (\mbox{h.c.}) \, \nonumber
\end{align}
where
\begin{align}
&c_{g}^{\,\alpha}({\bm p})g_{\bm p}(x) = C_f({\bm p})\big(2ip^\alpha - \partial^\alpha\big)g_{\bm p}(x) \return
&\bar{C}_{g}({\bm p})g_{\bm p}(x) = \bar{c}_f^{\,\alpha}({\bm p})\big(ip_\alpha - 2\partial_\alpha\big)g_{\bm p}(x) \,.
\end{align}

There is actually some freedom with how one defines these ghost decompositions based solely on the EOMs, but this freedom is fixed by requiring consistency with the BRST transformations of the operators shown below.
\begin{align}
\begin{aligned}
&\bigCom{\Chg}{h_f^{\alpha\beta}} = p^\alpha c_f^{\,\beta} + p^\beta c_f^{\,\alpha} - i\immet C_f \return
&\bigCom{\Chg}{H_f^{\alpha\beta}} = p^\alpha C_f^{\beta} + p^\beta C_f^{\,\alpha} + 4i p^\alpha p^\beta C_f \return
&\bigCom{\Chg}{A_f^\alpha} = iC_f^\alpha
\end{aligned}
\end{align}
\vspace*{-1.2em}
\begin{align}
\begin{aligned}
&\bigCom{\Chg}{b_f^\alpha} = 0 \qquad\qquad &&\bigCom{\Chg}{B_f^\alpha} = 0 \qquad\qquad &&\bigCom{\Chg}{B_f} = 0 \return
&\bigPB{\Chg}{c_f^\alpha} = 0 &&\bigPB{\Chg}{C_f^\alpha} = 0 &&\bigPB{\Chg}{C_f} = 0 \return
&\bigPB{\Chg}{\bar{c}_f^\alpha} = b_f^\alpha &&\bigPB{\Chg}{\bar{C}_f^\alpha} = B_f^\alpha &&\bigPB{\Chg}{\bar{C}_f} = B_f
\end{aligned}
\end{align}
Note, it is often more convenient to express the BRST transformations of operators in terms of their (anti)commutators with the BRST charge operator $\Chg$ as we have done here. These two pictures are related by $\delta_\epsilon X=\epsilon\Com{i\Chg}{X}_\mp$ where $\mp$ stands for commutator or anti-commutator as appropriate.

Our next task is to establish the (anti)commutation relations between the asymptotic fields which are obtained directly from the propagator (\ref{Prop}) using the replacements
\begin{align} \label{deltaRules}
&\frac{p_\alpha}{-p^2} \to -i\partial^x_\alpha D(x-y) \quad &\frac{p_\alpha}{(-p^2)^2} \to -i\partial^x_\alpha E(x-y) \quad &&\frac{p_\alpha}{(-p^2)^3} \to -i\partial^x_\alpha F(x-y) \qcom
\end{align}
where $D(x-y)$, $E(x-y)$, and $F(x-y)$ are the invariant delta functions for the first, second, and third powers of the d'Alembertian respectively. These functions are constructed from sums of the plane wave functions over ${\bm p}$; their precise definitions and important properties are displayed in Appendix \ref{sec:planewaves}. Using (\ref{deltaRules}), the non-zero commutators in the bosonic sector are found to be the following, where the superscript $x$ indicates differentiation with respect to $x$ only.
\begin{align} \label{BoFieldComs}
\begin{aligned}
&\bigCom{h_{\alpha\beta}(x)}{h_{\gamma\delta}(y)} = \frac{1}{2}\bigg(\Big(\delta_{\alpha\beta\gamma\delta} - \frac{1}{2}\eta_{\alpha\beta}\eta_{\gamma\delta}\Big)E(x - y) - \mathcal{D}^x_{\alpha\beta\gamma\delta}F(x - y)\bigg) \return
&\bigCom{h_{\alpha\beta}(x)}{H_{\gamma\delta}(y)} = \Big(\delta_{\alpha\beta\gamma\delta} - \frac{1}{2}\eta_{\alpha\beta}\eta_{\gamma\delta}\Big)D(x - y) - \mathcal{D}^x_{\alpha\beta\gamma\delta}E(x - y) \return
&\bigCom{h_{\alpha\beta}(x)}{b_{\gamma}(y)} = \big(\eta_{\alpha\gamma}\partial^x_\beta + \eta_{\beta\gamma}\partial^x_\alpha\big)D(x-y) \return
&\bigCom{h_{\alpha\beta}(x)}{B(y)} = \eta_{\alpha\beta}D(x-y) + 2\partial^x_\alpha\partial^x_\beta E(x-y) \return
&\bigCom{H_{\alpha\beta}(x)}{B_{\gamma}(y)} = \big(\eta_{\alpha\gamma}\partial^x_\beta + \eta_{\beta\gamma}\partial^x_\alpha\big)D(x-y) \return
&\bigCom{A_{\alpha}(x)}{A_{\beta}(y)} = \eta_{\alpha\beta}D(x-y) \return
&\bigCom{A_{\alpha}(x)}{B_{\beta}(y)} = -\eta_{\alpha\beta}D(x-y) \return
&\bigCom{A_{\alpha}(x)}{B(y)} = 2\partial^x_\alpha D(x-y)
\end{aligned}
\end{align}
Similarly, we find the following non-zero anti-commutators in the ghost sector.
\begin{align} \label{GhFieldComs}
\begin{aligned}
&\bigPB{c_\alpha (x)}{\bar{c}_\beta (y)} = i\eta_{\alpha\beta}D(x-y) \qquad\qquad &&\bigPB{c_\alpha (x)}{\bar{C} (y)} = i\partial^x_\alpha E(x-y) \return
&\bigPB{C_\alpha (x)}{\bar{C}_\beta (y)} = i\eta_{\alpha\beta}D(x-y) &&\bigPB{C_\alpha (x)}{\bar{C}(y)} = -2i\partial^x_\alpha D(x-y) \return
&\bigPB{C(x)}{\bar{C}(y)} = iD(x-y)
\end{aligned}
\end{align}

Finally, we obtain the non-zero (anti)commutators between the oscillators in each sector, which are consistent with the oscillator decompositions and the field commutators above (in the continuum limit), and are uniquely determined as the coefficients of $-p^{-2}$ in the propagator (\ref{Prop}).
\begin{align} \label{BoComs}
\begin{aligned}
&\BigCom{\osc{h}{f}{}_{\alpha\beta}({\bm p})}{\oscd{H}{f}{}_{\gamma\delta}({\bm q})} = \Big(\delta_{\alpha\beta\gamma\delta} - \frac{1}{2}\eta_{\alpha\beta}\eta_{\gamma\delta}\Big)\delta^3({\bm p} - {\bm q}) \return
&\BigCom{\osc{h}{f}{}_{\alpha\beta}({\bm p})}{\oscd{b}{f}{}_\gamma({\bm q})} = \big(ip_\alpha\eta_{\beta\gamma} +ip_\beta\eta_{\alpha\gamma}\big)\delta^3({\bm p} - {\bm q}) \return
&\BigCom{\osc{h}{f}{}_{\alpha\beta}({\bm p})}{\oscd{B}{f}{}({\bm q})} = \mmet\delta^3({\bm p} - {\bm q}) \return
&\BigCom{\osc{H}{f}{}_{\alpha\beta}({\bm p})}{\oscd{B}{f}{}_{\gamma}({\bm q})} = \big(ip_\alpha\eta_{\beta\gamma} +ip_\beta\eta_{\alpha\gamma}\big)\delta^3({\bm p} - {\bm q})\return
&\BigCom{\osc{A}{f}{}_{\alpha}({\bm p})}{\oscd{A}{f}{}_\beta({\bm q})} = \mmet\delta^3({\bm p} - {\bm q}) \return
&\BigCom{\osc{A}{f}{}_{\alpha}({\bm p})}{\oscd{B}{f}{}_\beta({\bm q})} = -\mmet\delta^3({\bm p} - {\bm q}) \return
&\BigCom{\osc{A}{f}{}_{\alpha}({\bm p})}{\oscd{B}{f}{}({\bm q})} = 2ip_\alpha\delta^3({\bm p} - {\bm q})
\end{aligned}
\end{align}
\begin{align} \label{GhComs}
\begin{aligned}
&\BigPB{\osc{c}{f}{}_{\alpha}({\bm p})}{\oscd{\bar{c}}{f}{}_\beta({\bm q})} = i\mmet\delta^3({\bm p} - {\bm q}) \qquad &&\BigPB{\osc{C}{f}{}_{\alpha}({\bm p})}{\oscd{\bar{C}}{f}{}_\beta({\bm q})} = i\mmet\delta^3({\bm p} - {\bm q}) \return
&\BigPB{\osc{C}{f}{}_{\alpha}({\bm p})}{\oscd{\bar{C}}{f}{}({\bm q})} = 2p_\alpha\delta^3({\bm p} - {\bm q}) &&\BigPB{\osc{C}{f}{}({\bm p})}{\oscd{\bar{C}}{f}{}({\bm q})} = i\delta^3({\bm p} - {\bm q})
\end{aligned}
\end{align}
These relations define the quantum theory of SOCG, whose total Fock space is spanned by the creation and annihilation operators contained in each of the oscillators. With this we may proceed by establishing the physical subspace of states in line with the BRST formalism.

\section{Unitarity} \label{sec:unitarity}

\subsection{The quartet mechanism}

In the covariant BRST quantization, one may classify all of the quantum states in a theory into two distinct groups; BRST singlets, which are identified as physical states, and BRST quartets of unphysical states whose total contribution to any scattering amplitude always sums to zero. These quartets consist of pairs of ``parent'' states $\ket{\pi}$ and ``daughter'' states $\ket{\delta}$ that are related by BRST transformation as
\begin{align}
\ket{\delta_{g+1}} = \Chg\ket{\pi_g} \neq 0 \qcom
\end{align}
where the subscripts indicate FP ghost number. The precise way in which this cancellation of unphysical states occurs is known as the Kugo-Ojima quartet mechanism \cite{Kugo1978-1,Kugo1979a,Kugo1979}, the proof of which relies on showing that the following relationship between inner products of parents and daughters holds.
\begin{align} \label{qrel}
\braket{\pi_{-1}}{\delta_1} = \bra{\pi_{-1}}\Chg\ket{\pi_0} = \braket{\delta_0}{\pi_0} \neq 0
\end{align}
Demonstrating this equality explicitly is enough to guarantee that only physical transverse states will contribute to scattering amplitudes in a given theory (more details regarding this fact and the quartet mechanism are collected in Appendix \ref{sec:BRSToverview}).

In order to show this for SOCG, we must reparameterize the states defined by the oscillators in the previous section. For convenience, and without loss of generality \cite{Kugo1978-2}, we restrict ourselves to a Lorentz frame defined by motion along the $z$-axis as defined by
\begin{align} \label{zframe}
p^\alpha = \big\{E,\,0,\,0,\,E\big\} \qcom
\end{align}
recalling that all (anti)commutators derived in this frame are also valid in general. With this in mind, we use the transverse oscillator equations (\ref{transh}\,-\ref{transA}) to eliminate nine of the twenty four components of $\osc{h}{f}{}_{\alpha\beta}$, $\osc{H}{f}{}_{\alpha\beta}$, and $\osc{A}{f}{}_{\alpha}$. Using the remaining fifteen independent components, we then define the six operators
\begin{align}
&\aop{h}{\pm} = \frac{1}{2}\Big(\osc{h}{f}{}_{11} - \osc{h}{f}{}_{22}\Big) \mp i\osc{h}{f}{}_{12} \label{ahDef} \return
&\aop{H}{\pm} = \frac{1}{2}\Big(\osc{H}{f}{}_{11} - \osc{H}{f}{}_{22}\Big) \mp i\osc{H}{f}{}_{12} \label{aHDef} \return
&\aop{A}{\pm} = \frac{1}{\sqrt{2}}\bigg(\osc{A}{f}{}_1 - \frac{i\osc{H}{f}{}_{13}}{E} \mp i\bigg(\osc{A}{f}{}_2 - \frac{i\osc{H}{f}{}_{23}}{E}\bigg)\bigg) \label{aADef} \qcom
\end{align}
which are all BRST singlets.
\begin{align}
&\BigCom{\Chg}{\aop{h}{\pm}} = 0 &&\BigCom{\Chg}{\aop{H}{\pm}} = 0  &\BigCom{\Chg}{\aop{A}{\pm}} = 0 \label{optransphys}
\end{align}

While there are many valid ways to define BRST-invariant operators from the independent components of our original fields, these particular choices allow us to write
\begin{align}
&\osc{h}{f}{}_{\alpha\beta}(\bm p) = \polt{+}_{\alpha\beta}(\bm p)\aop{h}{+}({\bm p}) + \polt{-}_{\alpha\beta}(\bm p)\aop{h}{-}({\bm p}) + (\mbox{h.c.}) + \cdots \label{hpols} \return
&\osc{H}{f}{}_{\alpha\beta}(\bm p) = \polt{+}_{\alpha\beta}(\bm p)\aop{H}{+}({\bm p}) + \polt{-}_{\alpha\beta}(\bm p)\aop{H}{-}({\bm p}) + (\mbox{h.c.}) + \cdots \label{Hpols} \return
&\osc{A}{f}{}_{\alpha}(\bm p) = \polt{+}_{\alpha}(\bm p)\aop{A}{+}({\bm p}) + \polt{-}_{\alpha}(\bm p)\aop{A}{-}({\bm p}) + (\mbox{h.c.}) + \cdots
\end{align}
where $\polt{\pm}_{\alpha\beta}$ and $\polt{\pm}_{\alpha}$ are circular polarization tensors corresponding to transverse $\pm 2$ and $\pm 1$ helicities respectively (see Appendix \ref{sec:PolTensors} for more on these polarization tensors). We thus identify $\aop{h}{\pm}$, $\aop{H}{\pm}$, and $\aop{A}{\pm}$ as the annihilation operators of physically propagating spin-2 and spin-1 fields. 

The ``$\cdots$'' above stand for the all the remaining (unphysical) longitudinal parts of the fundamental oscillators whose nine independent annihilation operators are combined into the nine BRST-non-trivial operators below.
\begin{align} \label{newops1}
&\Big(\psi_\alpha\Big) = \Array{
	-\frac{i\osc{h}{f}{}_{00}}{2E} + \frac{i\osc{H}{f}{}_{33}}{8E^3} - \frac{\osc{A}{f}{}_{3}}{4E^2} \\
	-\frac{i\osc{h}{f}{}_{01}}{E} \\
	-\frac{i\osc{h}{f}{}_{02}}{E} \\
	\frac{i\osc{h}{f}{}_{33}}{2E} + \frac{i\osc{H}{f}{}_{33}}{8E^3} - \frac{\osc{A}{f}{}_{3}}{4E^2}} \quad
&&\Big(\Psi_\alpha\Big) = \Array{
	\osc{A}{f}{}_{0} \\
	-\frac{i\osc{H}{f}{}_{01}}{E} \\
	-\frac{i\osc{H}{f}{}_{02}}{E} \\
	\osc{A}{f}{}_{3}} \quad
&&\Psi = \frac{\osc{H}{f}{}_{00}}{4E^2} - \frac{i\osc{A}{f}{}_{0}}{2E}
\end{align}
Finally, we also redefine the scalar NL and anti-ghost annihilation operators as
\begin{align} \label{newops2}
&\mathcal{B} = \osc{B}{f} - 2iE\big(\osc{B}{f}{}_0 + \osc{B}{f}{}_3\big) &\bar{\mathcal{C}} = \osc{\bar{C}}{f} - 2iE\big(\osc{\bar{C}}{f}{}_0 + \osc{\bar{C}}{f}{}_3\big) \,.
\end{align}
One finds that these quartet participants transform under BRST as
\begin{align} \label{optransq} 
\begin{aligned}
&\BigCom{\Chg}{\psi_\alpha} = i\osc{c}{f}{}_{\alpha} \qquad\qquad &&\BigCom{\Chg}{\Psi_\alpha} = i\osc{C}{f}{}_{\alpha} \qquad\qquad &&\BigCom{\Chg}{\Psi} = i\osc{C}{f} \return
&\BigCom{\Chg}{\osc{b}{f}{}_\alpha} = 0 &&\BigCom{\Chg}{\osc{B}{f}{}_\alpha} = 0 &&\BigCom{\Chg}{\mathcal{B}} = 0 \return
&\BigPB{\Chg}{\osc{c}{f}{}_\alpha} = 0 &&\BigPB{\Chg}{\osc{C}{f}{}_\alpha} = 0 &&\BigPB{\Chg}{\osc{C}{f}} = 0 \return
&\BigPB{\Chg}{\osc{\bar{c}}{f}{}_{\alpha}} = \osc{b}{f}{}_\alpha &&\BigPB{\Chg}{\osc{\bar{C}}{f}{}_\alpha} = \osc{B}{f}{}_\alpha &&\BigPB{\Chg}{\bar{\mathcal{C}}} = \mathcal{B} \,.
\end{aligned}
\end{align}

Crucially, the map between all of the new operators and the independent components of the original oscillators is invertible, guaranteeing that they span the entire Fock space. With the definitions (\ref{newops1},\,\ref{newops2}), it is straightforward to derive the (anti)commutation relations among each of our new creation and annihilation operators using (\ref{BoComs},\,\ref{GhComs}). In the physical subspace we have
\begin{align} \label{oprelsphys}
&\BigCom{\aop{h}{\lambda}({\bm p})}{\aop{H}{\lambda'}^\dagger({\bm q})} = \delta_{\lambda\lambda'}\delta^3({\bm p} - {\bm q}) \qquad &&\BigCom{\aop{A}{\lambda}({\bm p})}{\aop{A}{\lambda'}^\dagger({\bm q})} = \delta_{\lambda\lambda'}\delta^3({\bm p} - {\bm q}) 
\end{align}
where $\lambda=\pm$, while in the quartet subspace we find
\begin{align} \label{oprelsq}
&\BigCom{\psi_\alpha({\bm p})}{\oscd{b}{f}{}_\beta({\bm q})} = -\eta_{\alpha\beta}\delta^3(\bm p - \bm q) \qquad &&\BigCom{\Psi_\alpha({\bm p})}{\oscd{B}{f}{}_\beta({\bm q})} = -\eta_{\alpha\beta}\delta^3(\bm p - \bm q) \return
&\BigCom{\Psi({\bm p})}{\mathcal{B}^\dagger({\bm q})} = -\delta^3(\bm p - \bm q) &&\BigPB{\osc{c}{f}{}_\alpha({\bm p})}{\oscd{\bar{c}}{f}{}_\beta({\bm q})} = i\eta_{\alpha\beta}\delta^3(\bm p - \bm q) \return
&\BigPB{\osc{C}{f}{}_\alpha({\bm p})}{\oscd{\bar{C}}{f}{}_\beta({\bm q})} = i\eta_{\alpha\beta}\delta^3(\bm p - \bm q) &&\BigPB{\osc{C}{f}({\bm p})}{\bar{\mathcal{C}}^\dagger({\bm q})} = i\delta^3({\bm p} - {\bm q}) 
\end{align}
All of the (anti)commutation relations not shown above are identically zero, with the exception of some of the commutators between $\psi_\alpha$, $\Psi_\alpha$, and $\Psi$ that are non-trivial functions of $\bm p$. We do not show any of those commutators here as their precise values are irrelevant with regard to the KO quartet mechanism \cite{Nakanishi1990}.

Now, using the transformation properties (\ref{optransphys},\,\ref{optransq}) and the (anti)commutation relations (\ref{oprelsphys},\,\ref{oprelsq}), it is clear that the $\aop{i}{\lambda}$ states fully determine the physical subspace in this theory, while all the remaining states may be classified into parent and daughter states as
\begin{align}
&\big\{\ket{\pi_{0}}\big\} = \Big\{\psi^\dagger_\alpha\ket{0},\,\Psi^\dagger_\alpha\ket{0},\,\Psi^\dagger\ket{0}\Big\} \return
&\big\{\ket{\delta_{1}}\big\} = \big\{\Chg\ket{\pi_{0}}\big\} = \Big\{i\oscd{c}{f}{}_\alpha\ket{0},\,i\oscd{C}{f}{}_\alpha\ket{0},\,i\oscd{C}{f}\ket{0}\Big\} \return
&\big\{\ket{\pi_{-1}}\big\} = \Big\{-\oscd{\bar{c}}{f}{}_\alpha\ket{0},\,-\oscd{\bar{C}}{f}{}_\alpha\ket{0},\,-\bar{\mathcal{C}}^\dagger\ket{0}\Big\} \return
&\big\{\ket{\delta_{0}}\big\} = \big\{\Chg\ket{\pi_{-1}}\big\} = \Big\{-\oscd{b}{f}{}_\alpha\ket{0},\,-\oscd{B}{f}{}_\alpha\ket{0},\,-\mathcal{B}^\dagger\ket{0}\Big\} \,,
\end{align}
which are confined to quartets characterized by the defining relation (\ref{qrel}). Thus, as outlined at the beginning of this section, all of the quartet states appear strictly in zero norm combinations and do not threaten the unitarity of the transverse physical Fock space.

\subsection{Analysis of the physical subspace}

To investigate unitarity in the physical subspace of our theory, we will  follow the approach used by Kugo and Ojima to demonstrate the unitarity of Yang-Mills theories in \cite{Kugo1978-1,Kugo1979a,Kugo1979}. A key feature of these theories is that they come equipped with a positive definite inner product on their transverse physical Fock space $\FS_\text{tr}$,
\begin{align}
\braket{f}{f} > 0 \quad\forall\quad \ket{f} \in \FS_\text{tr} \qcom \ket{f} \neq 0 \qper
\end{align}
In fact, this relationship paired with a pseudo-unitary S-matrix that leaves the vacuum invariant and commutes with both the total (gauge-fixed) Hamiltonian and the BRST charge,
\begin{gather} \label{pseudouni}
\SM\SM^\dagger=\SM^\dagger\SM = \mathbbm{1} \qquad\qquad \SM\ket{0} = \SM^\dagger\ket{0} = \ket{0} \return
\bigCom{\Ham}{\SM} = \bigCom{\Chg}{\SM} = 0 \,,
\end{gather}
is enough to show the unitarity of a theory on very general grounds. Indeed, it is generally accepted that these three requirements may be used as the very definition of a unitary quantum theory.

The theory of conformal gravity we have presented here represents a special case however, as the non-diagonal nature of the commutation relations (\ref{oprelsphys}) implies the existence of an indefinite inner product on a portion of $\FS_\text{tr}$.  This fact alone does not imply a violation of unitarity however, only that we must investigate the issue with care. There is in fact a large amount of literature on the subject of indefinite metric QFTs; Nakanishi's work \cite{Nakanishi1972} being particularly helpful for what follows. However, before getting into the details of this investigation, we can make our task a bit easier by noting that a distinct portion of our $\mathcal{V}_\text{tr}$ already meets the unitarity requirements above, namely, the subspace spanned only by the $\aop{A}{\lambda}$ (with $\lambda=\pm$) where
\begin{align}
\mel{0}{\aop{A}{\lambda}({\bm p})\aop{A}{\lambda'}^\dagger({\bm q})}{0} = \delta_{\lambda\lambda'}\delta^3({\bm p} - {\bm q}) \qper
\end{align}
Thus, for the purposes of this discussion, we will restrict ourselves to the subspace of $\mathcal{V}_\text{tr}$ whose unitarity is in question i.e.\ that spanned only by $\aop{h}{\lambda}$ and $\aop{H}{\lambda}$, since the subspace involving $\aop{A}{\lambda}$ is easily seen to be unitary on its own and does not interact with the rest of $\mathcal{V}_\text{tr}$.

\subsubsection{Hamiltonian eigenstates}

To investigate unitarity in the remaining physical subspace, we begin by constructing its (gauge-fixed) Hamiltonian operator $\Ham$ using the Heisenberg equation
\begin{align}
\bigCom{\Ham}{\phi(x)} = -i\partial_0\phi(x) \qcom
\end{align}
where $\phi(x)=\{h_{\alpha\beta}(x),\,H_{\alpha\beta}(x)\}$. The right side of this equation is determined using the relations
\begin{align}
&i\partial_0f_{\bm p}(x) = Ef_{\bm p}(x) &&i\partial_0g_{\bm p}(x) = Eg_{\bm p}(x) + \frac{1}{2E}f_{\bm p}(x) \qper
\end{align}
The precise form of the Hamiltonian operator, which contains only creation and annihilation operators for spin-2 transverse polarizations, may thus be inferred by looking at the Heisenberg equation for each $\phi(x)$.
\begin{align} \label{Hamphys}
\Ham = \int\dd^3{\bm p}\sum_{\lambda=\pm}\left(E\Big(\aop{h}{\lambda}^\dagger(\bm p)\aop{H}{\lambda}(\bm p) + \aop{H}{\lambda}^\dagger(\bm p)\aop{h}{\lambda}(\bm p)\Big) + \frac{1}{4E}\Big(\aop{H}{\lambda}^\dagger(\bm p)\aop{H}{\lambda}(\bm p)\Big)\right)
\end{align}
This Hamiltonian is normal-ordered with respect to the vacuum as defined by
\begin{align}
\aop{h}{\lambda}(\bm p)\ket{0} = \aop{H}{\lambda}(\bm p)\ket{0} = 0 \qcom
\end{align}
and it commutes with the state operators according to the following relations.
\begin{align} \label{HaComs}
&\bigCom{\Ham}{\aop{h}{\lambda}^\dag(\bm p)} = E\aop{h}{\lambda}^\dag(\bm p) + \frac{1}{4E}\aop{H}{\lambda}(\bm p) \qquad\qquad \bigCom{\Ham}{\aop{H}{\lambda}^\dag(\bm p)} = E\aop{H}{\lambda}^\dag(\bm p)
\end{align}
We denote one-particle eigenstates of $\Ham$ as $\ket{{\bm p},\lambda}$, which may be written as a general linear combination of creation operators acting on the vacuum.
\begin{align}
\ket{{\bm p},\lambda} = \big(k_h\aop{h}{\lambda}^\dagger(\bm p) + k_H \aop{H}{\lambda}^\dagger(\bm p)\big)\ket{0}
\end{align}
Application of the Hamiltonian (\ref{Hamphys}) on this general state yields the eigenvalue equation
\begin{align}
\Ham\ket{{\bm p},\lambda} = E\left(k_h\aop{h}{\lambda}^\dagger(\bm p) + \left(k_H + \frac{k_h}{4E^2}\right)\aop{H}{\lambda}^\dagger(\bm p)\right)\ket{0} = \lambda_E\ket{{\bm p},\lambda} \qcom
\end{align}
from which we see that $k_h=0$ and $\lambda_E=E$. Thus, there exists only one one-particle energy eigenstate. The generalization to $n$-particle states is given by
\begin{gather} \label{nstateH}
\ket{{\bm p_n},\lambda_n}_H \equiv \frac{1}{\sqrt{n!}}\aop{H}{\lambda_1}^\dagger(\bm p_1)\cdots\aop{H}{\lambda_n}^\dagger(\bm p_n)\ket{0} \return
\Ham\ket{{\bm p_n},\lambda_n}_H = \sum_{i=1}^n\big(E_i\big)\ket{{\bm p_n},\lambda_n}_H \,,
\end{gather}
though this is only a subset of the more general complete set of eigenstates.

There is no one-particle (ket) eigenstate corresponding to $\aop{h}{\lambda}^\dag$ since the last term in (\ref{Hamphys}) effectively converts $\aop{h}{\lambda}^\dag$ to $\aop{H}{\lambda}^\dag$, however, it is possible to construct a particular type of multiparticle eigenstate using both $\aop{h}{\lambda}^\dag$ and $\aop{H}{\lambda}^\dag$ in such a way that the necessary cancellations occur\footnote{We thank Taichiro Kugo for pointing out the existence of this additional type of eigenstate.}. To demonstrate this explicitly, we consider a general state containing at least one $\aop{h}{\lambda}^\dag$,
\begin{align} \label{genhstate}
\cdots\aop{h}{\lambda}^\dag(\bm p)\cdots\ket{0} \,, 
\end{align}
where the $\cdots$ stands for an arbitrary products of $\aop{h}{\lambda}^\dag$ and $\aop{H}{\lambda}^\dag$. If we act on this state with $\Ham$, the $\aop{h}{\lambda}^\dag(\bm p)$ is converted to $(4E_{\bm p})^{-1}\aop{H}{\lambda}^\dag({\bm p})$, which must be canceled if we are to have an eigenstate. Pulling out an a $\aop{H}{\lambda}^\dag(\bm q)$ from the arbitrary $\cdots$ in (\ref{genhstate}) to write
\begin{align}
\cdots E_{\bm p}\aop{h}{\lambda}^\dag(\bm p)\aop{H}{\lambda}^\dag(\bm q)\cdots\ket{0} \,,
\end{align}
we see that the undesired conversion of $\aop{h}{\lambda}^\dag(\bm p)$ may be canceled by the analogous conversion that comes from a state of the form
\begin{align}
(-1)\cdots E_{\bm q}\aop{h}{\lambda}^\dag(\bm q)\aop{H}{\lambda}^\dag(\bm p)\cdots\ket{0} \,.
\end{align}
Therefore, the (normalized) linear combination of $\aop{h}{\lambda}^\dag$ and $\aop{H}{\lambda}^\dag$ given by
\begin{align} \label{Astate}
A_\lambda^\dag(\bm p,\bm q) = \frac{1}{2}\Big(\big(E_{\bm p}/E_{\bm q}\big)^{1/2}\aop{h}{\lambda}^\dag(\bm p)\aop{H}{\lambda}^\dag(\bm q) - \big(E_{\bm q}/E_{\bm p}\big)^{1/2}\aop{h}{\lambda}^\dag(\bm q)\aop{H}{\lambda}^\dag(\bm p)\Big)
\end{align}
is a unique eigenstate with the eigenvalue $(E_{\bm p}+E_{\bm q})$. Using the commutation relations (\ref{oprelsphys}), the norm of this new state is found to be
\begin{align} \label{Anorm}
\bra{0}A_\lambda({\bm p},{\bm q} )A^\dag_{\lambda'}({\bm p'},{\bm q'})\ket{0} = \delta_{\lambda\lambda'}D({\bm p},{\bm q};{\bm p'},{\bm q'}) \,,
\end{align}
where
\begin{align}
D({\bm p},{\bm q};{\bm p'},{\bm q'}) = \frac{1}{2}\big(\delta^3({\bm p} - {\bm q'})\delta^3({\bm q} - {\bm p'}) - \delta^3({\bm p} - {\bm p'})\delta^3({\bm q} - {\bm q'})\big) \,. 
\end{align}
Taking this additional type of eigenstate into account, we may write the most general $m,n$-particle eigenstate of $\Ham$ as
\begin{align}
&\ket{\bm p_m,\bm q_m,\lambda_m;\bm k_n,\zeta_n}_H \nreturn
&\qquad\equiv \frac{1}{(m!\,n!)^{1/2}}A_{\lambda_1}^\dag({\bm p}_1,{\bm q}_1) \cdots A_{\lambda_m}^\dag(\bm p_m,\bm q_m)\aop{H}{\zeta_1}^\dag(\bm k_1) \cdots \aop{H}{\zeta_n}^\dag(\bm k_n)\ket{0} \,, \label{genestate}
\end{align}
which has the eigenvalue $\sum_{i=1}^m (E_{{\bm p}_i}+E_{{\bm q}_i})+\sum_{j=1}^n E_{{\bm k}_j}$. 

We must also consider that, due to the vanishing commutation relation between $\aop{H}{\lambda}$ and $\aop{H}{\lambda}^\dagger$, the scalar product of (\ref{nstateH}) with its dual $\prescript{}{H}{\bra{\bm p_n,\lambda_n}}$ is zero. This in turn means that the scalar product between (\ref{genestate}) and itself also vanishes. The commutation relations (\ref{BoComs}) indicate that the only non-vanishing scalar product on the physical spin-2 subspace is between (\ref{genestate}) and the state
\begin{align} \label{genhstate}
&\prescript{}{h}{\bra{\bm p'_m,\bm q'_m,\lambda'_m;\bm k'_n,\zeta'_n}} \nreturn
&\qquad\equiv \frac{1}{(m!\,n!)^{1/2}}\bra{0}A_{\lambda'_1}(\bm p'_1,\bm q'_1) \cdots A_{\lambda'_m}(\bm p'_m,\bm q'_m)\aop{h}{\zeta'_1}(\bm k'_1) \cdots \aop{h}{\zeta'_n}(\bm k'_n) \,,
\end{align}
which is simply the bra version of (\ref{genestate}) with $H \leftrightarrow h$, as indicated by the prescript $h$. The subspaces spanned by (\ref{genestate}) and (\ref{genhstate}) respectively are obviously isomorphic and the only non-vanishing scalar product involving general eigenstates is thus given by 
\begin{align}
&\tensor[_h]{\braket{\bm p'_m,\bm q'_m,\lambda'_m;\bm k'_n,\zeta'_n}{\bm p_m,\bm q_m,\lambda_m;\bm k_n,\zeta_n}}{_H} \nreturn
&\qquad =\frac{1}{m!\,n!} \Big[\big(\delta_{\lambda'_1\lambda_1} \cdots \delta_{\lambda'_m\lambda_m}D(\bm p'_1,\bm q'_1;\bm p_1,\bm q_1) \cdots D(\bm p'_m,\bm q'_m;\bm p_m,\bm q_m) + \mbox{permutations} \big) \nreturn
&\phantom{\qquad =}\,\,\times\big(\delta_{\zeta'_1\zeta_1} \cdots \delta_{\zeta'_n\zeta_n}\delta^3(\bm k'_1 - \bm k_1) \cdots \delta^3(\bm k'_n - \bm k_n)
+ \mbox{permutations}\big)\Big] \,, \label{gennorm}
\end{align}
which in turn allows us to express the unit operator on the space of the $\Ham$ eigenstates as
\begin{align}
\mathbbm{1} = \sum_{m,n=0}\sum_{\lambda_m,\zeta_n}\int\,\dd^3\bm p_m\dd^3\bm q_m\dd^3\bm k_n (-1)^m\ket{\bm p_m,\bm q_m,\lambda_m;\bm k_n,\zeta_n}_H\prescript{}{h}{\bra{\bm p_m,\bm q_m,\lambda_m;\bm k_n,\zeta_n}} \,, \label{unitop1}
\end{align}
where the $m,n=0$ term should be understood as the vacuum contribution. Note that the factor of $(-1)^m$ that appears in this expression comes from the eigenstate (\ref{Astate}), whose existence is a consequence of the indefinite inner-product metric. The multi-particle eigenstates (\ref{genestate}) form a complete basis on the space of the $\Ham$ eigenstates; we denote this space as $\FS_\Ham$.

\subsubsection{The LSZ reduction formula and physical matrix elements} \label{sec:LSZ}

Having established the Hamiltonian operator and an understanding of the corresponding eigensystem, we may finally begin to investigate the defining characteristics of our S-matrix and how it relates to unitarity of the theory at large. Under the LSZ formalism, one assumes asymptotic completeness as defined by
\begin{align}
\FS^{\text{in}} \simeq \FS^{\text{out}} \simeq \FS \qand \SM\FS = \SM^\dagger\FS = \FS \,,
\end{align}
and as we saw in the previous section, $\FS_\Ham\subset\FS_\text{tr}$ is the only subspace that contains energy eigenstates (after also setting aside the obviously unitary subspace of spin-1 states). This is an important feature because, since $\FS_\Ham^{\text{in}}$ and $\FS_\Ham^{\text{out}}$ are Hamiltonian-invariant subspaces, we may use the pseudo-unitarity of $S$ to write
\begin{align}
\SM\FS_\Ham^{\text{in(out)}} = \FS_\Ham^{\text{in(out)}} \label{unitary1}\qcom
\end{align}
provided that the Hamiltonian also commutes with $S$. This then implies that the unitarity of $\SM$ on $\FS_\Ham^{\text{in}}$ may be expressed by either of the equivalent requirements below.
\begin{align} \label{UniReqmat}
\SM^\dagger\FS_\Ham^{\text{in}} = \FS_\Ham^{\text{in}} \quad\Leftrightarrow\quad \SM^\dagger\mathbbm{1}\SM = \SM\mathbbm{1}\SM^\dagger = \mathbbm{1}
\end{align}
Recalling the definition of the unit operator (\ref{unitop1}) on ${\cal V}_\Ham$, we restore the suppressed ``in'' and ``out'' designations on our creation and annihilation operators ($\aop{h}{\lambda} \rightarrow \aop{h}{\lambda}^\text{in}$, etc.) and consider
\begin{align} 
&\ket{\bm p_m,\bm q_m,\lambda_m;\bm k_n,\zeta_n;\text{in}}_H \nreturn
&\quad = \mathbbm{1}^\text{out}\ket{\bm p_m,\bm q_m,\lambda_m;\bm k_n,\zeta_n;\text{in}}_H \return
&\quad = \sum_{m',n'=0}\sum_{\lambda'_{m'},\zeta'_{n'}}\int\dd^3\bm p'_{m'}\dd^3\bm q'_{m'}\dd^3\bm k'_{n'}(-1)^{m'}\ket{\bm p'_{m'},\bm q'_{m'},\lambda'_{m'};\bm k'_{n'},\zeta'_{n'};\text{out}}_H S_{\beta\alpha} \,, \nonumber
\end{align}  
where $\mathbbm{1}^\text{out}$ is simply (\ref{unitop1}) written purely in terms of ``out'' states and
\begin{align}
S_{\beta\alpha} = \tensor[_h]{\braket{\beta;\text{out}}{\alpha;\text{in}}}{_H}
\end{align}
is the physical matrix element given by the transition amplitude between the physical states labeled by $\alpha$ and $\beta$ which are in turn defined as
\begin{align} \label{abdefs}
&\alpha = \big(\bm p_m,\bm q_m,\lambda_m;\bm k_n,\zeta_n\big) &&\beta = \big(\bm p'_{m'},\bm q'_{m'},\lambda'_{m'};\bm k'_{n'},\zeta'_{n'}\big) \,.
\end{align}

We can express the matrix elements $S_{\beta\alpha}$ in terms of time-ordered correlation functions by means of the LSZ reduction formula, which requires us to invert the oscillator decompositions and express our creation and annihilation operators in terms of the (physical) Heisenberg fields. Using (\ref{hdecomp},\,\ref{Hdecomp}) and (\ref{hpols},\,\ref{Hpols}), these fields may be fully decomposed as
\begin{align}
&h^\text{as}_{\alpha\beta}(x) = \sum_{\bm p,\lambda}\Big(\polt{\lambda}_{\alpha\beta}(\bm p)\aop{h}{\lambda}^\text{as}(\bm p)f_{\bm p}(x) + \frac{1}{2}\polt{\lambda}_{\alpha\beta}(\bm p)\aop{H}{\lambda}^\text{as}(\bm p)g_{\bm p}(x) + \text{(h.c.)} + \cdots\Big) \return
&H^\text{as}_{\alpha\beta}(x) = \sum_{\bm p,\lambda}\Big(\polt{\lambda}_{\alpha\beta}(\bm p)\aop{H}{\lambda}^\text{as}(\bm p)f_{\bm p}(x) + \text{(h.c.)} + \cdots\Big) \,,
\end{align}
where the ``$\cdots$'' represent the unphysical contributions from the quartet constituents as well as contributions from $\aop{A}{\lambda}(\bm p)$ which may all be neglected since these terms do not contribute to transverse $H$-$h$ scattering events. Using (\ref{PhihatDef}) and that $\poltd{\lambda}_{\alpha\beta}\polt{\lambda'}^{\alpha\beta}=\delta_{\lambda\lambda'}$, the inverses of these decompositions are found to be\footnote{A similar situation appears in the system of  the NL field and the would-be Goldstone boson in spontaneously broken Yang-Mills theory \cite{Kugo1979a}.}
\begin{align}
&\aop{h}{\lambda}^\text{as}(\bm p) = i\int\dd^3\bm x\,\poltd{\lambda}^{\alpha\beta}(\bm p)\Big(f^*_{\bm p}\overset{\leftrightarrow}{\partial_0}h^\text{as}_{\alpha\beta}(x) + g^*_{\bm p}\overset{\leftrightarrow}{\partial_0}\Box h^\text{as}_{\alpha\beta}(x)\Big) 
\label{ahDefhH} \return
&\aop{H}{\lambda}^\text{as}(\bm p) = i\int\dd^3\bm x\,
\poltd{\lambda}^{\alpha\beta}(\bm p)f^*_{\bm p}\overset{\leftrightarrow}{\partial_0}
H^\text{as}_{\alpha\beta}(x) \,. \label{aHDefH}
\end{align}
We then appeal to the fundamental theorem of calculus,
\begin{align}
\Big(\lim_{t\rightarrow\infty} - \lim_{t\rightarrow-\infty}\Big)\int\dd^3\bm x F(x) = \int^\infty_{-\infty}\dd x^0 \frac{\partial}{\partial x^0}\int\dd^3\bm x F(x) = \int\dd^4x\partial_0F(x) \qcom
\end{align}
and find that the in and out states overlap as follows, where we have used the relations $\partial_0^2f_{\bm p}=\nabla^2f_{\bm p}$ and $\partial_0^2
g_{\bm p}=\nabla^2g_{\bm p} - f_{\bm p}$ to write all derivatives in terms of $\Box$.
\begin{align}
&\aop{H}{\lambda}^{\text{out}\,\dagger}(\bm p)T(\cdots) - T(\cdots)\aop{H}{\lambda}^{\text{in}\,\dagger}(\bm p) = i\int\dd^4x\,\polt{\lambda}_{\alpha\beta}(\bm p)f_{\bm p}(x)\Box\,T\big(H^{\alpha\beta}(x)\cdots\big) \return
&\aop{h}{\lambda}^\text{out}(\bm p)T(\cdots)  - T(\cdots)\aop{h}{\lambda}^\text{in}(\bm p) = -\,i\int\dd^4x\,\poltd{\lambda}_{\alpha\beta}(\bm p)\Big(f^*_{\bm p}(x)\Box \,T\big(h^{\alpha\beta}(x) \cdots\big) \nreturn 
&\hspace{5.8cm} + \big(g^*_{\bm p}(x)\Box - f^*_{\bm p}(x)\big)T\big(\Box h^{\alpha\beta}(x) \cdots\big)\Big)\,,
\label{ahoverlap} 
\end{align}
where $T$ is a time-ordered product of the fields $(\cdots)$. Now, assuming that $\Box$ commutes with the $T$ product and using the graviton equation of motion
\begin{align} \label{hEOM}
\Box \grav(x) = \frac{1}{2}\Hab(x) + \cdots \,,
\end{align}
where the ``$\cdots$'' represent longitudinal terms that vanish thanks to the transverse polarization tensors as well as $\Ord(\alpha_g)$ interaction terms that also vanish in the reduction formula due to their lack of poles, we may simplify the $\aop{h}{\lambda}$ overlap above and write\footnote{The reduction formula (\ref{LSZrf}) will pick up additional terms if $\Box$ does not commute with $T$ product in (\ref{ahoverlap}).}
\begin{align}
\aop{h}{\lambda}^\text{out}(\bm p)T(\cdots) - T(\cdots)\aop{h}{\lambda}^\text{in}(\bm p) = -\frac{i}{2}\int\dd^4x\,\poltd{\lambda}_{\alpha\beta}(\bm p)g^*_{\bm p}(x)\Box\,T\big(H^{\alpha\beta}(x)\cdots\big) \,.
\end{align}
With the  preparations above we can arrive at the LSZ reduction formula by the standard construction \cite{Lehmann1955}, also taking care to account for disconnected terms in line with the cluster decomposition principle \cite{Weinberg1996}.
\begin{align} \label{LSZrf}
&\tensor[_h]{\braket{\bm p'_{m'},\bm q'_{m'},\lambda'_{m'};\bm k'_{n'},\zeta'_{n'};\text{out}}{\bm p_m,\bm q_m,\lambda_m;\bm k_n,\zeta_n;\text{in}}}{_H} \nreturn
&\qquad = \prod_{k=1}^{m'}\bigg[\!-\frac{1}{4}\int\dd^4x'_k\dd^4y'_k\Big(\big(E_{\bm p'_k}/E_{\bm q'_k}\big)^{1/2}\poltd{\lambda'_k}^{\!\!\alpha'_k\beta'_k}(\bm p'_k)\poltd{\lambda'_k}^{\!\!\gamma'_k\delta'_k}(\bm q'_k)
g^*_{\bm p'_k}(x'_k)f^*_{\bm q'_k}(y'_k) \nreturn
&\qquad\quad\,\, - \big({\bm p'}_k \leftrightarrow {\bm q'}_k\big)\Big)\Box_{x'_k}\Box_{y'_k}\bigg]\prod_{l=1}^{n'}\bigg[\!-\frac{i}{2}\int\dd^4z'_l\,\poltd{\zeta'_l}^{\!\!\mu'_l\nu'_l}(\bm k'_l)g^*_{\bm k'_l}(z'_l)\Box_{z'_l}\bigg] \nreturn
&\qquad\quad\,\prod_{i=1}^m\bigg[\!-\frac{1}{4}\int\dd^4x_i\dd^4y_i\Big(\,\big(E_{\bm p_i}/E_{\bm q_i}\big)^{1/2}\polt{\lambda_i}^{\!\!\alpha_i\beta_i}(\bm p_i)\polt{\lambda_i}^{\!\!\gamma_i\delta_i}(\bm q_i)g_{\bm p_i}(x_i)f_{\bm q_i}(y_i) \nreturn
&\qquad\quad\,\, - \big({\bm p}_i \leftrightarrow {\bm q}_i\big)\Big)\Box_{x_i}\Box_{y_i}\bigg]\prod_{j=1}^n\bigg[\!-i\int\dd^4z_j\,\polt{\zeta_j}^{\!\!\mu_j\nu_j}(\bm k_j)f_{\bm k_j}(z_j)\Box_{z_j}\bigg]\nreturn
&\qquad\quad\,\, \times G_{\alpha'_1\cdots\nu_n}\big(x'_1,\cdots,z_n\big) \,,
\end{align}
where 
\begin{align}
&G_{\alpha'_1\cdots\nu_n}\big(x'_1,\cdots,z_n\big) \nreturn
&\quad=\bra{0}TH_{\alpha'_1\beta'_1}(x'_1) \cdots H_{\alpha'_{m'}\beta_{m'}}(x'_{m'}) H_{\gamma'_1\delta'_1}(y'_1) \cdots H_{\gamma'_{m'}\delta'_{m'}}(y'_{m'})H_{\mu'_1\nu'_1}(z'_1) \cdots H_{\mu'_{n'}\nu'_{n'}}(z'_{n'}) \nreturn
&\quad\qquad\,\, H_{\alpha_1\beta_1}(x_1) \cdots H_{\alpha_{m}\beta_{m}}(x_{m}) H_{\gamma_1\delta_1}(y_1) \cdots H_{\gamma_{m}\delta_{m}}(y_{m}) H_{\mu_1\nu_1}(z_1) \cdots H_{\mu_{n}\nu_{n}}(z_{n})\ket{0} \,.
\end{align}
We note that, though it is beyond the scope of this work, we can in principle compute these Greens functions and hence the matrix element (\ref{LSZrf}) in perturbation theory at any finite order using the formalism developed in Sec. \ref{sec:classical} and \ref{sec:quant}. 

\subsubsection{Failure of the probability interpretation} \label{sec:ProbInt}

Using the LSZ reduction formula for the matrix elements that we have just derived, we are finally in a position to see exactly how the ghost problem in quadratic gravity manifests itself in the present formalism. In short, one finds that because of the indefinite metric structure (\ref{unitop1}) and the associated atypical Hamiltonian eigenstate (\ref{Astate}), $|S_{\beta\alpha}|^2$ can not be interpreted as the 
probability for some initial state $\alpha$ to transit to some final state $\beta$ in the traditional sense. To see this explicitly, we recall (\ref{unitary1}) and consider
\begin{align}  \label{UniReq}
1 = \tensor[_h]{\braket{\alpha;\text{in}}{\alpha;\text{in}}}{_H} = \tensor[_h]{\mel{\alpha;\text{in}}{\mathbbm{1}}{\alpha;\text{in}}}{_H} = \tensor[_h]{\mel{\alpha;\text{in}}{\SM^\dagger\mathbbm{1}\SM}{\alpha;\text{in}}}{_H}
\qcom
\end{align}
which, using the unit operator $\mathbbm{1}$ given in (\ref{unitop1}), we can write as
\begin{align}
1 = &\sum_{m,n=0}\sum_{\lambda_m,\zeta_n}\int\dd^3\bm p_m\dd^3\bm q_m\dd^3\bm k_n\Big( \nreturn
&(-1)^m\prescript{}{h}{\bra{\alpha;\text{in}}}S^\dag\ket{\bm p_m,\bm q_m,\lambda_m;\bm k_n,\zeta_n;\text{in}}_H\prescript{}{h}{\bra{\bm p_m,\bm q_m,\lambda_m;\bm k_n,\zeta_n;\text{in}}}S\ket{\alpha;\text{in}}_H\Big) \,. \label{SdagS}
\end{align}
Because of the presence of $(-1)^m$, the integrand of (\ref{SdagS}) is clearly not positive definite which implies, for general $n$ and $m$, 
\begin{align}
\prescript{}{h}{\bra{\alpha;\text{in}}}S^\dag\ket{\bm p_m,\bm q_m,\lambda_m;\bm k_n,\zeta_n;\text{in}}_H \neq \prescript{}{H}{\bra{\alpha;\text{in}}}S^\dag\ket{\bm p_m,\bm q_m,\lambda_m;\bm k_n,\zeta_n}_h \,.
\end{align}
Therefore, we can not interpret the integrand of (\ref{SdagS}) as the probability for an initial state $\alpha$ to transition to the $(2m+n)$-particle state $\beta$ (defined similarly to (\ref{abdefs})), as we should be able to in a unitary theory. Although this result is expected, to our knowledge this is the first time an explicit demonstration of the violation of unitarity in Weyl's conformal gravity has been identified in the canonical covariant operator formalism.

\section{Conclusion}

The second order formulation of conformal gravity presented here makes the covariant operator formalism for BRST quantization possible and consequently, allows the LSZ formalism to be applied in the traditional fashion. The motivation behind this work was an in-depth investigation of the ghost problem in scale-symmetric quadratic gravity and we have thus put our focus on the quadratic (in $h_{\alpha\beta}$) part of the  linearized action. Though higher order interaction terms are of course important in general, it is the kinetic terms that determine whether the theory is unitary or not, provided that one assumes the LSZ formalism with interacting Heisenberg fields that behave asymptotically as free fields when $t\to\pm\infty$.

To arrive at a second order formulation of the classical theory, we introduced the auxiliary tensor field $H_{\alpha\beta}$. This step alone is enough to allow for an application of the Dirac-Bergman algorithm to count classical degrees of freedom in the Hamiltonian formalism. If one introduces $H_{\alpha\beta}$ alone, one finds ten first class and eight second class constraints for a total of $1/2(2 \times 20 - 2 \times 10 - 8) = 6$ degrees of freedom in configuration space, as expected from other classical studies of conformal gravity \cite{Riegert1984}. However, one is also always free to redefine phase-space variables in order to exchange second with first class constraints, which at the level of the action, is equivalent to our introduction of the Stückelberg vector $A_\alpha$ and the accompanying additional diffeomorphism symmetry. With this reformulation one finds $1/2(2 \times 24 - 2 \times 18) = 6$ degrees of freedom as before, but in a system with purely first class constraints. 

This first class formulation not only makes the ensuing calculations simpler, but also makes it easier to classify the physical DOF according to their representation of the Poincaré group. The irreducible representations of the Poincaré group are of course fixed, and in four space-time dimensions, massless bosons with spin greater than zero have two physical polarizations. This implies that covariant vector and tensor fields necessarily carry unphysical DOF if they should describe massless particles. These unphysical DOF endanger the unitarity of any given theory, however, in the BRST formalism employed here, the unphysical DOF and the Faddeev-Popov ghosts \cite{Faddeev1969,Faddeev1975} form Kugo-Ojima quartets. The KO quartet mechanism\cite{Kugo1978-1,Kugo1979a,Kugo1979} ensures that $\mathcal{V}_\text{tr} \equiv \text{Ker}\,\Chg/\text{Im}\,\Chg $ (where $\Chg$ is the BRST charge \cite{Becchi1975,Becchi1976}) defines a physical subspace with no unphysical DOF (provided that no BRST anomalies are present). Interestingly, we are never forced to make any a priori assumptions about the polarizations of the physical modes in the present theory. Rather, we find that the transverse spin-2 and spin-1 representations were essentially selected for us by the KO quartet mechanism, based solely on the behavior of the fundamental oscillators under BRST transformation. Using this formalism we can in principle compute covariant correlation functions at any finite order in perturbation theory. We emphasize that the intermediate contributions to the covariant correlation functions coming from the above-mentioned 18 bosonic DOF are {\em exactly canceled} by the Faddeev-Popov ghosts.

For more traditional theories (Yang-Mills for example), the discussion on unitarity basically ends here since, if $\mathcal{V}_\text{tr}$ is a space with positive-definite metric, unitarity of the theory follows immediately \cite{Kugo1978-1,Kugo1979a,Kugo1979}. The situation is not so straight-forward for the present theory however. Though the spin-1 sector possesses a positive-definite metric, the metric in the spin-2 subspace is indefinite, a fact that appears to be a general feature of quantum gravitation \cite{Stelle1978,Kugo1978-2}. However, the indefinite metric turns out to be a remarkable feature since, though these transverse spin-2 states belong to the same class under the BRST cohomology (singlets), the gauge-fixed Hamiltonian distinguishes between them. There actually only exists one type of Hamiltonian eigenstate that forms a Hamiltonian invariant subspace i.e.\ a BRST invariant subspace that remains unchanged after scattering if the scattering operator commutes with the Hamiltonian. However, we have explicitly shown that the physical matrix elements (the transition amplitudes between two physical states) 
\begin{align}
S_{\beta\alpha} = \tensor[_h]{\mel{\beta;\text{in}}{S}{\alpha;\text{in}}}{_H} \,,
\end{align}
do not allow for an interpretation of probability, indicating that unitarity of the S-matrix on the physical subspace is violated. This fact should be interpreted as the way in which the ghost problem manifests itself in the covariant operator formalism.

For the case of globally scale symmetric quadratic gravity the story is much the same; the key differences being a lack of scalar ghost and NL fields, and two additional physical scalar modes. One of these scalars corresponds to the auxiliary scalaron field $\chi(x)$ while the other comes from the longitudinal component of $A_\alpha(x)$, which we may refer to as $\phi(x)$. While this longitudinal mode ends up as part of a KO quartet in the conformal case, it ends up propagating when there are no scalar ghosts and NL fields to partner with it. One then finds an additional independent physical subspace spanned by $a_\chi(\bm p)$ and $a_\phi(\bm p)$, the annihilation operators for the scalaron and longitudinal scalar respectively. These operators have no non-vanishing commutation relations with any of the other physical mode operators and they exhibit the exact same indefinite metric structure among themselves as each of the spin-2 polarizations,
\begin{align}
&\tensor[_\phi]{\braket{\beta;\text{in}}{\alpha;\text{in}}}{_\phi} = \tensor[_\chi]{\braket{\beta;\text{in}}{\alpha;\text{in}}}{_\chi} = 0 &\tensor[_\phi]{\braket{\beta;\text{in}}{\alpha;\text{in}}}{_\chi} = \delta_{\alpha\beta} \,.
\end{align}
Given this structure, an investigation of unitarity in the scalar sector follows in essentially the same, yet simpler way as for the spin-2 sector shown in the main text.

The present work will also serve as a starting point for a long list of future projects. First among these will be an in-depth study of the Weyl anomaly \cite{Capper1974,Capper1975,Deser1976,Casarin2018} which, under our scheme of covariant operator quantization will manifest as a BRST anomaly. We expect that the Wess-Zumino condition \cite{Wess1971,Bardeen1984} for the Weyl anomaly may be reformulated as a problem of BRST cohomology at ghost number unity \cite{Bonora1983}, making regularization independent analysis of the Weyl anomaly possible in our framework. An explicit calculation of the BRST anomaly will then likely show the consistency of our formalism with known results (see e.g.\ \cite{Birrell1994}), however it is also possible that, due to our novel second order method of quantization (in particular the presence of ghosts corresponding to the Weyl symmetry), interesting differences may appear that can be compared with other studies of the Weyl anomaly in conformal gravity \cite{Capper1975,Salvio2018a}.

Next, with our BRST symmetry in hand, it will be interesting to attempt to make regularization-independent statements about the renormalizability of quadratic gravity \cite{Pottel2020a}. Using the Zinn-Justin equation, which controls the structure of divergent terms \cite{Zinn-Justin1975}, a proof of renormalizability becomes yet another problem of BRST cohomology.
\\

\textbf{Acknowledgments:} We wish to thank Taichiro Kugo, Manfred Lindner, Jonas Rezacek, Philipp Saake, Klaus Sibold, Andreas Trautner, and Masatoshi Yamada for many helpful discussions. J. Kubo is partially supported by the Grant-in-Aid for Scientific Research (C) from the Japan Society for Promotion of Science (Grant No.19K03844).

\begin{appendix}

\section{Overview of BRST Quantization} \label{sec:BRSToverview}

To establish a consistent unitary quantum theory from a free gauge theory, it is important to account for the inherent over-counting of physically equivalent states that are related by gauge transformations. However, introducing gauge fixing terms into the action alone is not always enough to define a physical Fock space (particularly in the presence of non-simple gauge groups) and ensure that the resulting quantum theory will be renormalizable and stable to all orders in perturbation theory. In general one must establish a BRST symmetry in the theory and identify the physical subspace of interest with the space of states that are annihilated by the BRST charge operator \cite{Becchi1975,Becchi1976,Kugo1978-1}.

To set up the BRST construction in a given classical theory, one must first establish a gauge fixing action $S_{\text{gf}}$ by choosing a set of gauge fixing conditions $\chi^a$ that are functions of the fields $\phi^A$ present in the classical action $S_\text{cl}$, then introduce a set of bosonic auxiliary ``Nakanishi-Lautrup'' (NL) fields $B^a$ to act as Lagrange multipliers that enforce the $\chi^a$.
\begin{align} \label{Sgf}
S_{\text{gf}} = \int\dd^4x B_a\chi^a
\end{align}
The next step is to introduce additional fields that compensate for the unphysical states. These compensators are the Faddeev-Popov (FP) ghost and anti-ghost fields (\cite{Faddeev1969,Faddeev1975}) $C^a$ and $\bar{C}^a$, which are both Hermitian and independent of each other. FP (anti)ghosts carry integer spin, but obey fermionic (Grassmann) statistics which allows them to cancel with bosonic degrees of freedom. To ensure that only unphysical bosonic modes are compensated for, and to ensure that this cancellation is precise i.e.\ ensure that no spin-statistic-theorem-violating (anti)ghosts appear as asymptotic states, the (anti)ghosts must be introduced into the action so as to establish a BRST symmetry.

The defining object of a BRST symmetry is the fermionic charge $\Chg$ which generates the graded BRST algebra. This grading is known as the ``ghost number'' and one assigns
\begin{align}
&\text{gh}\big(\phi^A\big) = \text{gh}\big(B^a\big) = 0 &&\text{gh}\big(C^a\big) = 1 &&\text{gh}\big(\bar{C}^a\big) = -1 \qcom
\end{align}
while requiring that the action and all observables carry ghost number zero. $\Chg$ generates the following transformations of the fields
\begin{align} \label{genQtrans}
\begin{aligned}
&\delta_{\epsilon}\phi^A = \epsilon\,\sum_a\Big(\delta_{\xi^a}\phi^A\Big)\evltd{\xi^a=C^a} \qquad\qquad\qquad &&\delta_{\epsilon}B^a = 0 \return
&\delta_{\epsilon}C^a = \epsilon C^b\partial_bC^a &&\delta_{\epsilon}\bar{C}^a = \epsilon\,iB^a \qcom
\end{aligned}
\end{align}
where $\epsilon$ is a constant anti-commuting and anti-Hermitian parameter, and $\delta_{\xi^a}\phi^A$ is the gauge transformation of $\phi^A$ with the gauge parameter $\xi^a(x)$. Crucially, $\Chg$ is nilpotent ($\Chg^2=0$) which is what allows one to construct the BRST-invariant gauge-fixed total action $S_\text{T}$ by simply subtracting the quantity $\Chg(\bar{C}_a\chi^a)$ from the classical action $S_\text{cl}$, which is already BRST invariant as a result of its gauge symmetry. The charge operator's action on $\bar{C}_a\chi^a$ generates the desired gauge fixing terms, identical to (\ref{Sgf}), as well as the appropriate corresponding Faddeev-Popov ghost terms. One thus arrives at
\begin{align} \label{genST}
S_\text{T} = S_\text{cl} - \Chg\big(\bar{C}_a\chi^a\big) = S_\text{cl} + S_\text{gf} + S_\text{FP} \qcom\qquad \delta_{\epsilon}S_\text{T} = 0 \qper
\end{align}

Kugo and Ojima showed in \cite{Kugo1978-1,Kugo1979a,Kugo1979} that if an action possesses a BRST symmetry as described above, then the total Fock space $\FS\ni\ket{f}$, which generally contains unphysical states of negative norm, may be restricted to the physical subspace $\FSp\ni\ket{f_{\text{phys}}}$ via the simple requirement
\begin{align} \label{subcon}
\mathcal{Q}\ket{f_{\text{phys}}} = 0 \qper
\end{align}
From this requirement one then further identifies the subspace containing only physically propagating transverse states i.e.\ the space with no longitudinal or zero-norm states, as the quotient space\footnote{This quotient space is referred to as the ``BRST cohomology space''. The subject of BRST cohomology is a rich subject that we will not go into further detail on here, though we recommend \cite{Barnich2000,Dragon2012} (among many others) as resources for the curious reader.}
\begin{align}
\mathcal{V}_\text{tr} = \text{Ker}\,\Chg/\text{Im}\,\Chg \qcom
\end{align}
where $\text{Ker}\,\Chg = \FSp$ and $\text{Im}\,\Chg=\Chg\FS=\mathcal{V}_0$ is the BRST co-boundary \cite{Fujikawa1978}.

In practice, once a complete Fock space is established in a given theory one need only appeal to the ``Kugo-Ojima quartet mechanism'' to identify the precise transverse physical subspace. What follows here is a condensed version of the full proof of this mechanism laid out by Nakanishi in \cite{Nakanishi1990}. The KO mechanism functions by first noting that all the members of $\FS$ may be classified as singlets or doublets under BRST. We refer to members of the doublet as parent ($\ket{\pi}$) and daughter ($\ket{\delta}$) states, and they are characterized by the relation
\begin{align}
\ket{\delta_1} = \Chg\ket{\pi_0} \neq 0 \qcom
\end{align}
where the subscripts indicate FP ghost number. The nilpotency of $\Chg$ guarantees that no higher $n$-plet representations may appear. Furthermore, as a result of the requirement that the ghost number of the total action be zero, each doublet is necessarily accompanied by an FP-conjugate doublet $\{\ket{\pi_{-1}},\,\ket{\delta_0}\}$. These pairs of doublets live in the BRST co-boundary and constitute the titular quartets, while all the remaining BRST singlets fully populate $\mathcal{V}_\text{tr}$. Matrix elements between physical and quartet states are always vanishing, while non-vanishing matrix elements between members of the quartets always appear as the following.
\begin{align} \label{quartetDef}
\braket{\pi_{-1}}{\delta_1} = \bra{\pi_{-1}}\Chg\ket{\pi_0} = \braket{\delta_0}{\pi_0} \neq 0
\end{align}
This fact is guaranteed by taking advantage of the inherent freedom in defining the parent and daughter states. It is in fact always possible to define their familial relationship so that not only are the quartet states orthogonal to the physical states, but also that the $n$-particle states constructed from them may be written in terms of the projection operator
\begin{align}
P^{(n)} = \frac{1}{n}\sum_{a,b=1}^4\big(g^{-1})_{ab}\varphi_a^\dagger P^{(n-1)} \varphi_b \qcom
\end{align}
where and $g_{ab}$ is the inner product metric defined by $\Com{\varphi_a(\bm p)}{\varphi_b^\dagger(\bm q)}=g_{ab}\delta^3(\bm p - \bm q)$ and $\varphi_a = \{\pi_0,\,\delta_0,\,\pi_{-1},\,\delta_1\}$ are the annihilation operators for each of the quartet participants. Using this recursive relation and the definition $\mathcal{V}_\text{tr}=P^{(0)}\FS=$ (the subspace with zero unphysical particles), a straightforward proof by induction leads to the conclusion $\mathcal{V}_0=\sum_{n=1}^\infty P^{(n)}\FSp=\Chg\FS$, and hence
\begin{align}
\mel{f_{\text{phys}}}{P^{(n)}}{g_{\text{phys}}} = 0 \qcom n \geq 1 \qper
\end{align} 
Thus, even though the quartets may appear as asymptotic multi-particle states in $\FSp$ as defined by (\ref{subcon}), their scalar product with any state in $\mathcal{V}_\text{phys}$ is necessarily zero. In other words, states that compose part of a quartet as defined by (\ref{quartetDef}) do not pose a threat when establishing the unitarity of the physical S-matrix on $\mathcal{V}_\text{tr}$.

\section{Plane waves and Greens functions} \label{sec:planewaves}

Since dipole and tripole contributions to the propagators appear in our formulation of conformal gravity, we briefly summarize the definitions and properties of the plane wave solutions, invariant delta functions, and propagators (Greens functions) introduced in Section \ref{sec:AsymFields} as part of the LSZ formalism. This construction closely follows that found in \cite{Nakanishi1990}, though we have extended the formalism up through quadrupole functions since such terms appear in second order quadratic gravity as well as in the second order conformal theory if a different gauge is considered.

The plane wave solution ($\Box\,f_{\bm p}(x)=0$), normalized in a finite volume $V$, is denoted by
\begin{align}
f_{\bm p}(x) = \frac{1}{\sqrt{2EV}}e^{i p x} \,,
\end{align}
where $E=|{\bm p}|$ and $px=-E\,x^0+{\bm p}\cdot{\bm x}$. The plane wave solutions for dipole, tripole, and quadrupoles, denoted by $g_{\bm p}(x)$, $h_{\bm p}(x)$, and $k_{\bm p}(x)$ respectively, are given by
\begin{align}
&g_{\bm p}(x) = -\frac{1}{2\sqrt{2EV}}\left(\frac{1}{2E^2} + \frac{ix^0}{E}\right)e^{ipx} \label{gpDef}\return
&h_{\bm p}(x) = \frac{1}{8\sqrt{2EV}}\left(\frac{5}{4E^4} + \frac{2ix^0}{E^3} - \frac{(x^0)^2}{E^2}\right)e^{ipx} \return
&k_{\bm p}(x) = -\frac{1}{128\sqrt{2EV}}\left(\frac{45}{3E^6} + \frac{22ix^0}{E^5} - \frac{12(x^0)^2}{E^4} - \frac{8i(x^0)^3}{3E^3}\right)e^{ipx} \qcom
\end{align}
which satisfy
\begin{align}
&\Box^4k_{\bm p}(x) = 0 &&\Box\,k_{\bm p}(x) = h_{\bm p}(x) \return
&\Box^3h_{\bm p}(x) = 0 &&\Box\,h_{\bm p}(x) = g_{\bm p}(x) \return
&\Box^2g_{\bm p}(x) = 0 &&\Box\,g_{\bm p}(x) = f_{\bm p}(x) \qper
\end{align}
These solutions also satisfy the following orthogonality relations.
\begin{align}
&\int \dd^3{\bm x} \,f_{\bm p}^*(x)\overleftrightarrow{\partial_0}f_{\bm q}(x) = -i\delta_{{\bm p, \bm q}} \label{ortho1} \return
&\int \dd^3{\bm x} \left(f_{\bm p}^*(x)\overleftrightarrow{\partial_0}g_{\bm q}(x) + g_{\bm p}^*(x)\overleftrightarrow{\partial_0}f_{\bm q}(x)\right) = 0 \return
&\int \dd^3{\bm x} \left(f_{\bm p}^*(x)\overleftrightarrow{\partial_0}h_{\bm q}(x) + h_{\bm p}^*(x)\overleftrightarrow{\partial_0}f_{\bm q}(x) + g_{\bm p}^*(x)\overleftrightarrow{\partial_0}g_{\bm q}(x)\right) = 0 \return
&\int \dd^3{\bm x} \left(f_{\bm p}^*(x)\overleftrightarrow{\partial_0}k_{\bm q}(x) + k_{\bm p}^*(x)\overleftrightarrow{\partial_0}f_{\bm q}(x) + g_{\bm p}^*(x)\overleftrightarrow{\partial_0}h_{\bm q}(x) + h_{\bm p}^*(x)\overleftrightarrow{\partial_0}g_{\bm q}(x)\right) = 0 \label{ortho4}
\end{align}

Corresponding to the plane wave solutions $f_{\bm p}(x),\,g_{\bm p}(x) $ and $h_{\bm p}(x)$, we introduce the following invariant delta functions with positive frequency.
\begin{align}
D^{(+)}(x-y) &= \sum_{\bm p}f_{\bm p}(x)f_{\bm p}^*(y) = \sum_{\bm p}\frac{1}{2EV}\,e^{ip(x-y)} \return
E^{(+)}(x-y) &= \sum_{\bm p}\left(f_{\bm p}(x)g_{\bm p}^*(y) + g_{\bm p}(x)f_{\bm p}^*(y)\right) \nreturn
&= -\sum_{\bm p}\frac{1}{4EV}\left(\frac{1}{E^2} + \frac{i}{E}(x^0-y^0)\right)e^{ip(x-y)} \return
F^{(+)}(x-y) &= \sum_{\bm p}\left(f_{\bm p}(x)h_{\bm p}^*(y) + h_{\bm p}(x)f_{\bm p}^*(y) + g_{\bm p}(x)g_{\bm p}^*(y)\right) \nreturn
&= \sum_{\bm p}\frac{1}{16EV}\left(\frac{3}{E^4} + i\frac{3}{E^3}(x^0-y^0) - \frac{(x^0-y^0)^2}{E^2}\right)e^{i p (x-y)} \return
G^{(+)}(x-y) &= \sum_{\bm p}\left(f_{\bm p}(x)k_{\bm p}^*(y) + k_{\bm p}(x)f_{\bm p}^*(y) + g_{\bm p}(x)h_{\bm p}^*(y) + h_{\bm p}(x)g_{\bm p}^*(y)\right) \nreturn
&= \sum_{\bm p}\frac{1}{64EV}\left(-\frac{5}{E^6} - i\frac{5}{E^5}(x^0-y^0) + \frac{2(x^0-y^0)^2}{E^4} + \frac{i(x^0-y^0)^3}{3E^3}\right)e^{i p (x-y)}
\end{align}
These functions serve as Greens functions for the first, second, and third powers of the d'Alembertian.
\begin{align}
\begin{aligned}
&\Box\,D^{(+)}(x-y) = 0 &&\Box\,E^{(+)}(x-y) = D^{(+)}(x-y) \return
&\Box\,F^{(+)}(x-y) = E^{(+)}(x-y) \qquad\qquad &&\Box\,G^{(+)}(x-y) = F^{(+)}(x-y)
\end{aligned}
\end{align}
The invariant delta functions that appear in the commutator relations (\ref{BoFieldComs},\,\ref{GhFieldComs}) are constructed from the these functions and their negative frequency counterparts as
\begin{align}
\begin{aligned}
&D = D^{(+)} - D^{(-)} \qquad\qquad&&E = E^{(+)} - E^{(-)} \return
&F= F^{(+)} - F^{(-)} &&G = G^{(+)} - G^{(-)} \qcom
\end{aligned}
\end{align}
where $D^{(-)}=(D^{(+)})^*$, and similarly for the others. In the continuum limit $V\to \infty$, the sum over the momentum becomes an integral in three dimensional momentum space and the invariant functions may be written as
\begin{align}
&D(x-y) = \int \dd^3{\bm p} \frac{1}{2E(2\pi)^3}\Big[e^{ip(x-y)} - \text{(h.c.)}\Big] \return
&E(x-y) = -\int \dd^3{\bm p} \frac{1}{4E(2\pi)^3}\left[\left(\frac{1}{E^2} + i\frac{x^0 - y^0}{E}\right)e^{ip(x-y)} - \text{(h.c.)}\right] \return
&F(x-y) = \int \dd^3{\bm p} \frac{1}{16E(2\pi)^3}\left[\left(\frac{3}{E^4} + i\frac{3(x^0 - y^0)}{E^3} - \frac{(x^0 - y^0)^2}{E^2}\right)e^{ip(x-y)} - \text{(h.c.)}\right] \return
&G(x-y) = -\int \dd^3{\bm p} \frac{1}{64E(2\pi)^3}\bigg[\Big(\frac{5}{E^6} + i\frac{5}{E^5}(x^0-y^0) \nreturn
&\phantom{G(x-y) = }- \frac{2(x^0-y^0)^2}{E^4} - \frac{i(x^0-y^0)^3}{3E^3}\Big)e^{i p (x-y)} - \text{(h.c.)}\bigg]\qper
\end{align} 
These in turn satisfy
\begin{align}
D(x)\evltd{x^0 = 0} &= 0 &&\partial_0D(x)\evltd{x^0 = 0} = -i\delta({\bm x}) \return
\partial_0^nE(x)\evltd{x^0 = 0} &= 0 \qquad (n=0,1,2)  &&\partial_0^3E(x)\evltd{x^0 = 0} = i\delta({\bm x}) \return
\partial_0^nF(x)\evltd{x^0 = 0} &= 0 \qquad (n=0,\dots,4)  &&\partial_0^5F(x)\evltd{x^0 = 0} = -i\delta({\bm x}) \return
\partial_0^nG(x)\evltd{x^0 = 0} &= 0 \qquad (n=0,\dots,6)  &&\partial_0^7G(x)\evltd{x^0 = 0} = i\delta({\bm x}) \qper
\end{align} 
We also note that the Feynman propagators may be expressed in terms of the invariant delta functions as follows.
\begin{align}
D_F(x) &= \theta(x^0)D^{(+)}(x) + \theta(-x^0)D^{(-)}(x) = -i\int \frac{\dd^4p}{(2\pi)^4}\frac{e^{ipx}}{p^2 - i\epsilon} \return
E_F(x) &= \theta(x^0)E^{(+)}(x) + \theta(-x^0)E^{(-)}(x) = i\int \frac{\dd^4p}{(2\pi)^4}\frac{e^{ipx}}{(p^2 - i\epsilon)^2} \return
F_F(x) &= \theta(x^0)F^{(+)}(x) + \theta(-x^0)F^{(-)}(x) = -i\int \frac{\dd^4p}{(2\pi)^4}\frac{e^{ipx}}{(p^2 - i\epsilon)^3} \return
G_F(x) &= \theta(x^0)G^{(+)}(x) + \theta(-x^0)G^{(-)}(x) = i\int \frac{\dd^4p}{(2\pi)^4}\frac{e^{ipx}}{(p^2 - i\epsilon)^4}
\end{align}

An important object related to this construction is the integro-differential operator
\begin{align} \label{invBox}
\mathcal{E}^{(\eta)} = -\frac{1}{2}\big(\nabla^2\big)^{-1}\big(x^0\partial_0 - \eta\big) \qcom
\end{align}
where $\eta$ is an arbitrary dimensionless constant, that acts as an inverse d'Alembertian specifically when it acts on solutions to the d'Alembert equation.
\begin{align}
\Box\mathcal{E}^{(\eta)}f_{\bm p}(x) = f_{\bm p}(x)
\end{align}
This operator also functions as an inverse d'Alembertian when applied to the propagators shown here, with the caveat that $\eta=1$ for the Feynman propagators specifically.
\begin{align}
&\Box\mathcal{E}^{(\eta)}D^{(\pm)}(x) = D^{(\pm)}(x) &&
\Box\mathcal{E}^{(1)}D_F(x) = D_F(x)
\end{align}

\section{Transverse polarization tensors} \label{sec:PolTensors}

In this appendix we give an overview of the system of transverse polarization tensors used to describe the massless spin-1 and spin-2 fields in this work. We begin by defining a complete set of four polarization vectors $\polt{i}_\alpha(\bm p)$ where $i=0,\cdots,3$, that is orthonormal and complete.
\begin{align}
&\polt{i}_\alpha(\bm p)\polt{j}^\alpha(\bm p) = \eta_{ij} &&\sum_{i=0}^3\polt{i}_\alpha(\bm p)\polt{i}_\beta(\bm p) = \delta_{\alpha\beta}
\end{align}
In the Lorentz frame defined by motion purely along the $z$-axis (\ref{zframe}), we may identify the transverse polarizations where $\polt{i}_\alpha p^\alpha=0$ as
\begin{align}
&\big(\polt{1}_\alpha\big) = \Array{0 \\ 1 \\ 0 \\ 0} &&\big(\polt{2}_\alpha\big) = \Array{0 \\ 0 \\ 1 \\ 0} \qper
\end{align}
It is then a simple matter to find that the following combinations of these vectors represent $\pm 1$ helicities.
\begin{align}
\polt{\pm}_\alpha = \frac{1}{\sqrt{2}}\big(\polt{1}_\alpha \pm i\polt{2}_\alpha\big)
\end{align}
This fact may be confirmed by applying a general rotation around the $z$-axis.
\begin{align}
R_\alpha^{\beta}(\theta)\polt{\pm}_\beta = e^{\pm i\theta}\polt{\pm}_\alpha \qwhere \Big(R_\alpha^{\beta}(\theta)\Big) = 
\begin{pmatrix}
1 & 0 & 0 & 0 \\
0 & \cos\theta & \sin\theta & 0 \\
0 & -\sin\theta & \cos\theta & 0 \\
0 & 0 & 0 & 1
\end{pmatrix}
\end{align}
Then, with a Clebsch-Gordon table in hand, one finds that the polarization tensors for helicity $\pm 2$ are given by
\begin{align}
\Big(\polt{\pm}_{\alpha\beta}\Big) = \Big(\polt{\pm}_\alpha\otimes\polt{\pm}_\beta\Big) = 
\frac{1}{2}\begin{pmatrix}
0 & 0 & 0 & 0 \\
0 & 1 & \pm i & 0 \\
0 & \pm i & -1 & 0 \\
0 & 0 & 0 & 0
\end{pmatrix} \quad\text{with}\quad R_\alpha^{\gamma}(\theta)R_\beta^{\delta}(\theta)\polt{\pm}_{\gamma\delta} = e^{\pm 2i\theta}\polt{\pm}_{\alpha\beta} \qper
\end{align}
An important fact regarding these polarization tensors is that though they depend on the three-momentum $\bm p$ in a general Lorentz frame, being transverse, they depend only on the direction of the three-momentum and not on its magnitude i.e.\ $\partial/\partial E\big(\polt{\pm}_{\alpha\beta}(\bm p)\big) = 0$.

\end{appendix}

\vspace{-0.5cm}

\bibliographystyle{JHEP}
\bibliography{library}

\providecommand{\href}[2]{#2}\begingroup\raggedright\begin{thebibliography}{10}

\bibitem{Stelle1977}
K.~S. Stelle, {\it {Renormalization of higher-derivative quantum gravity}},
  {\em Physical Review D} {\bf 16} (aug, 1977) 953--969.

\bibitem{Ostrogradsky1850}
M.~Ostrogradsky, {\it {M{\'{e}}moires sur les {\'{e}}quations
  diff{\'{e}}rentielles, relatives au probl{\`{e}}me des
  isop{\'{e}}rim{\`{e}}tres}},  {\em Mem. Acad. St. Petersbourg} {\bf 6}
  (1850), no.~4 385--517.

\bibitem{Woodard2015}
R.~P. Woodard, {\it {The Theorem of Ostrogradsky}},
  \href{http://www.arxiv.org/abs/1506.02210}{{\tt 1506.02210}}.

\bibitem{Horndeski1974}
G.~W. Horndeski, {\it {Second-order scalar-tensor field equations in a
  four-dimensional space}},  {\em International Journal of Theoretical Physics}
  {\bf 10} (1974), no.~6 363--384.

\bibitem{Langlois2016}
D.~Langlois and K.~Noui, {\it {Degenerate higher derivative theories beyond
  Horndeski: Evading the Ostrogradski instability}},  {\em Journal of Cosmology
  and Astroparticle Physics} {\bf 2016} (2016), no.~2 1--19,
  [\href{http://www.arxiv.org/abs/1510.06930}{{\tt 1510.06930}}].

\bibitem{Boulware1983}
D.~G. Boulware, G.~T. Horowitz, and A.~Strominger, {\it {Zero-Energy Theorem
  for Scale-Invariant Gravity}},  {\em Physical Review Letters} {\bf 50} (may,
  1983) 1726--1729.

\bibitem{Donoghue2019}
J.~F. Donoghue and G.~Menezes, {\it {Unitarity, stability, and loops of
  unstable ghosts}},  {\em Physical Review D} {\bf 100} (2019), no.~10
  [\href{http://www.arxiv.org/abs/1908.02416}{{\tt 1908.02416}}].

\bibitem{Donoghue2021}
J.~F. Donoghue and G.~Menezes, {\it {The Ostrogradsky instability can be
  overcome by quantum physics}},
  \href{http://www.arxiv.org/abs/2105.00898}{{\tt 2105.00898}}.

\bibitem{Anselmi2018}
D.~Anselmi, {\it {Fakeons And Lee-Wick Models}},
  \href{http://www.arxiv.org/abs/1801.00915}{{\tt 1801.00915}}.

\bibitem{Lee1969a}
T.~D. Lee and G.~C. Wick, {\it {Negative metric and the unitarity of the
  S-matrix}},  {\em Nuclear Physics, Section B} {\bf 9} (1969), no.~2 209--243.

\bibitem{Salvio2016}
A.~Salvio and A.~Strumia, {\it {Quantum mechanics of 4-derivative theories}},
  {\em European Physical Journal C} {\bf 76} (2016), no.~4
  [\href{http://www.arxiv.org/abs/1512.01237}{{\tt 1512.01237}}].

\bibitem{Bender2008}
C.~M. Bender and P.~D. Mannheim, {\it {No-ghost theorem for the fourth-order
  derivative pais-uhlenbeck oscillator model}},  {\em Physical Review Letters}
  {\bf 100} (2008), no.~11 1--4,
  [\href{http://www.arxiv.org/abs/0706.0207}{{\tt 0706.0207}}].

\bibitem{Mannheim2018}
P.~D. Mannheim, {\it {Unitarity of loop diagrams for the ghost-like
  propagator}},  {\em arXiv} (2018), no.~5 1--18,
  [\href{http://www.arxiv.org/abs/1801.03220}{{\tt 1801.03220}}].

\bibitem{Kugo1979b}
T.~Kugo and I.~Ojima, {\it {Local Covariant Operator Formalism of Non-Abelian
  Gauge Theories and Quark Confinement Problem}},  {\em Progress of Theoretical
  Physics Supplement} {\bf 66} (1979), no.~66 1--130.

\bibitem{Nakanishi1990}
N.~Nakanishi and I.~Ojima, {\em {Covariant Operator Formalism Of Gauge Theories
  And Quantum Gravity}}.
\newblock World Scientific, 1990.

\bibitem{Kugo}
T.~Kugo, {\em {Eichtheorie}}.
\newblock Springer Verlag, 1997.

\bibitem{Kugo1978-2}
T.~Kugo and I.~Ojima, {\it {Subsidiary conditions and physical S-matrix
  unitarity in indefinite-metric quantum gravitation theory}},  {\em Nuclear
  Physics B} {\bf 144} (nov, 1978) 234--252.

\bibitem{Kugo2022}
T.~Kugo, R.~Nakayama, and N.~Ohta, {\it {Covariant BRST Quantization of
  Unimodular Gravity I -- Formulation with antisymmetric tensor ghosts}},
  \href{http://www.arxiv.org/abs/2202.03626}{{\tt 2202.03626}}.

\bibitem{Lehmann1955}
H.~Lehmann, K.~Symanzik, and W.~Zimmermann, {\it {Zur Formulierung
  quantisierter Feldtheorien}},  {\em Il Nuovo Cimento} {\bf 1} (1955), no.~1
  205--225.

\bibitem{Martin-Garcia2008}
J.~M. Mart{\'{i}}n-Garc{\'{i}}a, {\it {xPerm: fast index canonicalization for
  tensor computer algebra}},  {\em Computer Physics Communications} {\bf 179}
  (2008), no.~8 597--603, [\href{http://www.arxiv.org/abs/0803.0862}{{\tt
  0803.0862}}].

\bibitem{Brizuela2009}
D.~Brizuela, J.~M. Mart{\'{i}}n-Garc{\'{i}}a, and G.~A. {Mena Marug{\'{a}}n},
  {\it {xPert: Computer algebra for metric perturbation theory}},  {\em General
  Relativity and Gravitation} {\bf 41} (2009), no.~10 2415--2431,
  [\href{http://www.arxiv.org/abs/0807.0824}{{\tt 0807.0824}}].

\bibitem{Nutma2014a}
T.~Nutma, {\it {XTras: A field-theory inspired xAct package for mathematica}},
  {\em Computer Physics Communications} {\bf 185} (2014), no.~6 1719--1738,
  [\href{http://www.arxiv.org/abs/1308.3493}{{\tt 1308.3493}}].

\bibitem{Frob2020}
M.~B. Fr{\"{o}}b, {\it {FieldsX -- An extension package for the xAct tensor
  computer algebra suite to include fermions, gauge fields and BRST
  cohomology}},  \href{http://www.arxiv.org/abs/2008.12422}{{\tt 2008.12422}}.

\bibitem{Alvarez-Gaume2016}
L.~Alvarez-Gaume, A.~Kehagias, C.~Kounnas, D.~L{\"{u}}st, and A.~Riotto, {\it
  {Aspects of quadratic gravity}},  {\em Fortschritte der Physik} {\bf 64}
  (2016), no.~2-3 176--189, [\href{http://www.arxiv.org/abs/1505.07657}{{\tt
  1505.07657}}].

\bibitem{Riegert1984}
R.~J. Riegert, {\it {The particle content of linearized conformal gravity}},
  {\em Physics Letters A} {\bf 105} (1984), no.~3 110--112.

\bibitem{Kugo1978-1}
T.~Kugo and I.~Ojima, {\it {Manifestly Covariant Canonical Formulation of the
  Yang-Mills Field Theories I: General Formalism}},  {\em Progress of
  Theoretical Physics} {\bf 60} (dec, 1978) 1869--1889.

\bibitem{Kugo1979a}
T.~Kugo and I.~Ojima, {\it {Manifestly Covariant Canonical Formulation of
  Yang-Mills Field Theories II: SU(2) Higgs-Kibble Model with Spontaneous
  Symmetry Breaking}},  {\em Progress of Theoretical Physics} {\bf 61} (jan,
  1979) 294--314.

\bibitem{Kugo1979}
T.~Kugo and I.~Ojima, {\it {Manifestly Covariant Canonical Formulation of
  Yang-Mills Field Theories III: Pure Yang-Mills Theories without Spontaneous
  Symmetry Breaking}},  {\em Progress of Theoretical Physics} {\bf 61} (feb,
  1979) 644--655.

\bibitem{Nakanishi1972}
N.~Nakanishi, {\it {Indefinite-Metric Quantum Field Theory}},  {\em Progress of
  Theoretical Physics} {\bf 51} (1972) 1--95.

\bibitem{Weinberg1996}
S.~Weinberg, {\em {The Quantum Theory of Fields: Volume I - Foundations}}.
\newblock Cambridge University Press, 1996.

\bibitem{Faddeev1969}
L.~D. Faddeev, {\it {The Feynman Integral for Singular Lagrangians}},  {\em
  Theor. Math. Phys.} {\bf 1} (1969), no.~1 1--13.

\bibitem{Faddeev1975}
L.~D. Faddeev, {\it {Introduction to Functional Methods}},  pp.~1--40, 1975.

\bibitem{Becchi1975}
C.~Becchi, A.~Rouet, and R.~Stora, {\it {Renormalization of the abelian
  Higgs-Kibble model}},  {\em Communications in Mathematical Physics} {\bf 42}
  (1975), no.~2 127--162.

\bibitem{Becchi1976}
C.~Becchi, A.~Rouet, and R.~Stora, {\it {Renormalization of Gauge Theories}},
  {\em Annals of Physics} {\bf 98} (1976) 287--321.

\bibitem{Stelle1978}
K.~S. Stelle, {\it {Classical gravity with higher derivatives}},  {\em General
  Relativity and Gravitation} {\bf 9} (apr, 1978) 353--371.

\bibitem{Capper1974}
D.~M. Capper and M.~J. Duff, {\it {Trace anomalies in dimensional
  regularization}},  {\em Il Nuovo Cimento A Series 11} {\bf 23} (1974), no.~1
  173--183.

\bibitem{Capper1975}
D.~M. Capper and M.~J. Duff, {\it {Conformal anomalies and the
  renormalizability problem in quantum gravity}},  {\em Physics Letters A} {\bf
  53} (1975), no.~5 361--362.

\bibitem{Deser1976}
S.~Deser, M.~Duff, and C.~Isham, {\it {Non-local conformal anomalies}},  {\em
  Nuclear Physics B} {\bf 111} (aug, 1976) 45--55.

\bibitem{Casarin2018}
L.~Casarin, H.~Godazgar, and H.~Nicolai, {\it {Conformal anomaly for
  non-conformal scalar fields}},  {\em Physics Letters, Section B: Nuclear,
  Elementary Particle and High-Energy Physics} {\bf 787} (2018) 94--99,
  [\href{http://www.arxiv.org/abs/1809.06681}{{\tt 1809.06681}}].

\bibitem{Wess1971}
J.~Wess and B.~Zumino, {\it {Consequences of anomalous ward identities}},  {\em
  Physics Letters B} {\bf 37} (1971), no.~1 95--97.

\bibitem{Bardeen1984}
W.~A. Bardeen and B.~Zumino, {\it {Consistent and Covariant Anomalies in Gauge
  and Gravitational Theories}},  {\em Nuclear Physics} {\bf 244} (1984)
  421--453.

\bibitem{Bonora1983}
L.~Bonora, P.~Cotta-Ramusino, and C.~Reina, {\it {Conformal anomaly and
  cohomology}},  {\em Physics Letters B} {\bf 126} (1983), no.~5 305--308.

\bibitem{Birrell1994}
N.~D. Birrell and P.~C.~W. Davies, {\em {Quantum Fields in Curved Space}}.
\newblock Cambridge University Press, feb, 1994.

\bibitem{Salvio2018a}
A.~Salvio and A.~Strumia, {\it {Agravity up to infinite energy}},  {\em
  European Physical Journal C} {\bf 78} (2018), no.~2
  [\href{http://www.arxiv.org/abs/1705.03896}{{\tt 1705.03896}}].

\bibitem{Pottel2020a}
S.~Pottel and K.~Sibold, {\it {Perturbative quantization of Einstein-Hilbert
  gravity embedded in a higher derivative model}},  {\em Physical Review D}
  {\bf 104} (oct, 2021) 086012,
  [\href{http://www.arxiv.org/abs/2012.11450}{{\tt 2012.11450}}].

\bibitem{Zinn-Justin1975}
J.~Zinn-Justin, {\it {Renormalization of gauge theories}},  {\em Lect. Notes
  Phys.} {\bf 37} (1975) 1--39.

\bibitem{Barnich2000}
G.~Barnich, F.~Brandt, and M.~Henneaux, {\it {Local BRST cohomology in gauge
  theories}},  {\em Physics Report} {\bf 338} (2000), no.~5 439--569,
  [\href{http://www.arxiv.org/abs/0002245}{{\tt 0002245}}].

\bibitem{Dragon2012}
N.~Dragon and F.~Brandt, {\it {BRST Symmetry and Cohomology}},  {\em Strings,
  Gauge Fields, and the Geometry Behind: The Legacy of Maximilian Kreuzer}
  (2012) 3--86, [\href{http://www.arxiv.org/abs/1205.3293}{{\tt 1205.3293}}].

\bibitem{Fujikawa1978}
K.~Fujikawa, {\it {On a Superfield Theoretical Treatment of the Higgs-Kibble
  Mechanism}},  {\em Progress of Theoretical Physics} {\bf 59} (1978), no.~6
  2045--2063.

\end{thebibliography}\endgroup

\end{document}